# Continuous gravitational waves in the lab: recovering audio signals with a table-top optical microphone

James W. Gardner,[1, 2, 3, *] Hannah Middleton,[3, 4, 5, †] Changrong Liu,[6, 5, ‡]
Andrew Melatos,[3, 5, §] Robin Evans,[6, 5] William Moran,[6] Deeksha Beniwal,[7, 8, 9]
Huy Tuong Cao,[7, 8, 9] Craig Ingram,[7, 8, 9] Daniel Brown,[7, 8, 9] and Sebastian Ng[7, 8, 9]

[1] *Centre for Gravitational Astrophysics, The Australian National University, Acton, ACT, 2601, Australia*
[2] *OzGrav-ANU, Australian Research Council Centre of Excellence for Gravitational Wave Discovery, The Australian National University, Acton, ACT, 2601, Australia*
[3] *School of Physics, University of Melbourne, Parkville, Victoria, 3010, Australia*
[4] *Centre for Astrophysics and Supercomputing, Swinburne University of Technology, Hawthorn, Victoria, 3122, Australia*
[5] *OzGrav-Melbourne, Australian Research Council Centre of Excellence for Gravitational Wave Discovery, Parkville, Victoria, 3010, Australia*
[6] *Department of Electrical and Electronic Engineering, University of Melbourne, Parkville, Victoria, 3010, Australia*
[7] *Department of Physics, The University of Adelaide, South Australia, 5005, Australia*
[8] *The Institute of Photonics and Advanced Sensing (IPAS), The University of Adelaide, South Australia, 5005, Australia*
[9] *OzGrav-Adelaide, Australian Research Council Centre of Excellence for Gravitational Wave Discovery, South Australia, 5005, Australia*

Gravitational-wave observatories around the world are searching for continuous waves: persistent signals from sources such as spinning neutron stars. These searches use sophisticated statistical techniques to look for weak signals in noisy data. In this paper, we demonstrate these techniques using a table-top model gravitational-wave detector: a Michelson interferometer where sound is used as an analog for gravitational waves. Using signal processing techniques from continuous-wave searches, we demonstrate the recovery of tones with constant and wandering frequencies. We also explore the use of the interferometer as a teaching tool for educators in physics and electrical engineering by using it as an "optical microphone" to capture music and speech. A range of filtering techniques used to recover signals from noisy data are detailed in the Supplementary Material. Here, we present highlights of our results using a combined notch plus Wiener filter and the statistical log minimum mean-square error (logMMSE) estimator. Using these techniques, we easily recover recordings of simple chords and drums, but complex music and speech are more challenging. This demonstration can be used by educators in undergraduate laboratories and can be adapted for communicating gravitational-wave and signal-processing topics to non-specialist audiences.

## I. INTRODUCTION

In 2015, the first detection of gravitational waves from the merger of two black holes marked a breakthrough in modern astrophysics and revealed a new means to observe the Universe.[1] Gravitational waves are a prediction of Albert Einstein's theory of General Relativity; they are disturbances in spacetime caused by the acceleration of asymmetric massive objects. The effect of gravitational waves is a change in lengths: a "stretching and squashing" of the distance between objects. Ground-based gravitational-wave observatories such as the Advanced Laser Interferometer Gravitational-wave Observatory (LIGO), Advanced Virgo, GEO600, and KAGRA use the interference of laser light to measure changes in distance. These observatories are extremely complex but are fundamentally based on the Michelson interferometer. Table-top interferometers are commonly used in undergraduate laboratory experiments and to demonstrate the science of gravitational-wave detection to non-specialist audiences.[2]

To date, the network of gravitational-wave observatories has observed short-duration transient signals from the mergers of binary black holes, binary neutron stars, and binaries consisting of a neutron star and a black hole.[3, 4] However, the network is also searching for continuous gravitational waves: persistent, periodic, near-monochromatic signals, which are yet to be detected. Rotating neutron stars are prime candidates for continuous-wave emission, especially those in low mass X-ray binaries (LMXB), where the neutron star is in orbit with a low mass stellar companion. The rotation frequency of the neutron star in an LMXB can wander over time due to variable accretion of matter (and hence angular momentum transfer) from the stellar companion.[5] Scorpius X-1 is a prime LMXB target for continuous-wave searches. Numerous searches, as yet unsuccessful, have been performed for Scorpius X-1 and other LMXBs (e.g., Ref. [6]).

* james.gardner@anu.edu.au
† hannah.middleton@unimelb.edu.au
‡ changrongl1@student.unimelb.edu.au
§ amelatos@unimelb.edu.au

In this paper, we use a table-top Michelson interferometer as a toy gravitational-wave detector designed to detect sound instead of gravitational waves. We then extend its use to an "optical microphone", using light to capture sound, and present a range of example analysis techniques for educators to use. As an undergraduate lab experiment, the apparatus can be used to teach topics ranging from continuous-wave detection and analysis to electronics, signal processing, and speech enhancement. It allows students in courses such as physics and electrical engineering to explore the response of an accessible, yet nontrivial, optomechanical system using a hierarchy of data analysis techniques of increasing complexity, including those used in the search for continuous waves in LIGO and Virgo data.[6, 7] This demonstration also has the potential to be used as an outreach tool alongside a range of other public engagement demonstrations and activities developed by gravitational-wave research groups around the world. These tools allow scientists to cater to the increased public and media interest in this field and explain gravitational-wave science to non-specialist audiences.

This paper is laid out as follows. In Section II, we detail the table-top interferometer design. In Section III, we demonstrate observing a single tone from a speaker. In Section IV, we observe a wandering frequency signal and analyze it using a hidden Markov model technique (the Viterbi algorithm) from continuous-wave searches. In Section V, we demonstrate capture and playback of complex audio, such as music and speech. This demonstration of an optical microphone serves as a more general exhibition of signal processing with a range of examples that can be used in the undergraduate laboratory (described in the Supplementary Material). We suggest avenues of future work in Section VI and draw conclusions in Section VII. Further reading and resources are provided in the Supplementary Material and we present the software and scripts used to produce this work in Appendix A.

## II. TABLE-TOP GRAVITATIONAL-WAVE SCIENCE

Gravitational-wave detectors such as LIGO and Virgo are large, complex experiments. However, their design is fundamentally based on the Michelson interferometer, an optical configuration commonly used in undergraduate laboratories. In a Michelson interferometer, laser light is split by a beamsplitter into two perpendicular arms, shown in Fig. 1. Mirrors at the end of each arm reflect the two beams back to the beamsplitter where they recombine to produce an interference pattern. The resulting interference pattern is dependent on the relative distance traveled by the beams. Current generation gravitational-wave observatories have kilometer-scale arms (with arm lengths 4 km and 3 km at LIGO and Virgo, respectively). They can measure minuscule changes in distance due to gravitational waves; e.g., the first detection of a binary black hole merger produced a strain of $10^{-21}$,[1] corresponding to a change in distance equal to a fraction of the diameter of a proton.

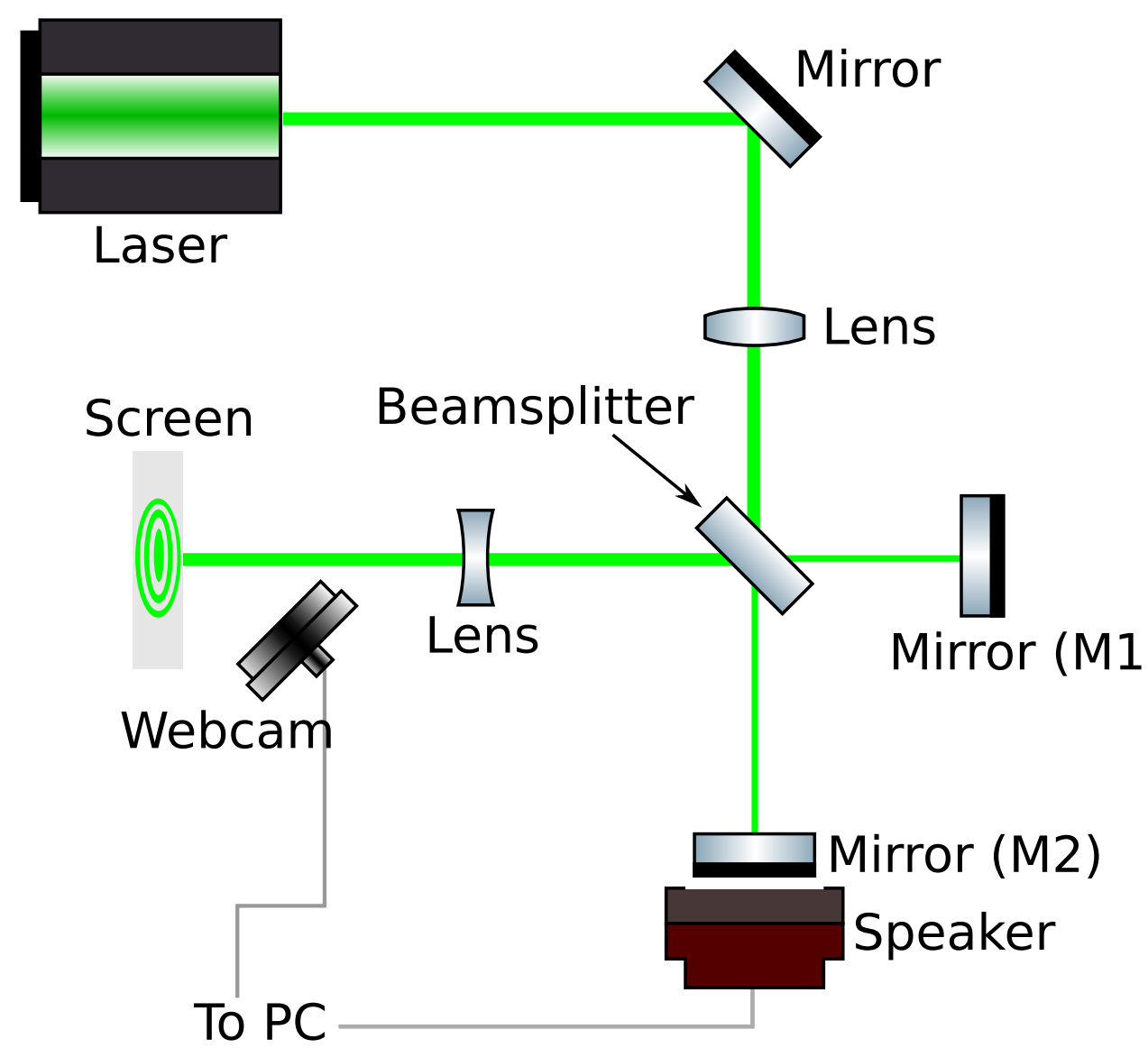

FIG. 1. Schematic of the table-top Michelson interferometer. The output of a Class 2 532 nm laser (top-left) is reflected from a mirror at a right angle. The beam then passes through a converging lens (focal length: 125.0 mm) before reaching the beamsplitter, which reflects approximately 50% of the beam and transmits the remaining $\sim 50\%$. The split beams are reflected from mirrors (M1 and M2) at the ends of the interferometer arms and recombine at the beamsplitter. The output interference pattern is enlarged using a diverging lens (focal length: $-25.0$ mm), projected onto a screen, and recorded using a commercial webcam connected to a computer (PC). A speaker fixed to the back of M2 is used to inject audio signals from the PC into the interferometer.

Sound is a commonly used analogy when explaining gravitational-wave science. Gravitational-wave signals from binary black hole and neutron star mergers can be converted to audio signals [8] to aid in explanations. Detection and analysis techniques used by the gravitational-wave community can be demonstrated using table-top equipment (see Ref. [2] as well as the further reading section in the Supplementary Material for a selection of other table-top Michelson interferometer designs used for science outreach). Sound vibrations provide a simple means to move the components of a demonstration interferometer, changing the length of the interferometer arms, and therefore the interference pattern. We use audio signals throughout this work to simulate gravitational wave–like signals.

The optical configuration of the Michelson interferometer used in this work is shown in Fig. 1. It is assembled on a $450 \times 450$ mm optical breadboard and uses a green laser with peak emission at a wavelength of 532 nm. The output of the laser is first reflected by a mirror which turns the beam 90° (to save space on the optical breadboard and to have greater control of the alignment of the interferometer). After passing through a converging (plano-convex) lens with focal length 125.0 mm, the laser beam is incident on a beamsplitter, which reflects half of the light towards mirror 1 (M1) and transmits the other half to mirror 2 (M2). A 0.5 W speaker is fixed to the back of M2 using commercial adhesive putty and serves as a controllable source of vibrations. This speaker is one of a pair of commercial, USB-powered speakers, fed by a 3.5 mm jack and driven by a computer (PC). The other speaker in the pair is kept face-down and as far away from the interferometer as possible to prevent it from acting as a second source. The beams are reflected by mirrors M1 and M2, located at the end of $\sim 7.5$ cm and $\sim 10.0$ cm long arms. The beams recombine at the beamsplitter and produce an interference pattern that is then enlarged using a diverging (bi-concave) lens of focal length $-25.0$ mm, and projected onto a screen. The interference pattern, as shown by the photograph in Fig. 2, is a set of concentric light and dark rings (fringes). A change in the relative length of the arms causes these rings to move radially inwards or outwards. The intensity timeseries is recorded by either a webcam (in Sections III and IV) or a photodiode, which offers a higher sampling rate suitable for capturing more complex audio (in Section V). Further design details for Michelson interferometers can be found in Ref. [2].

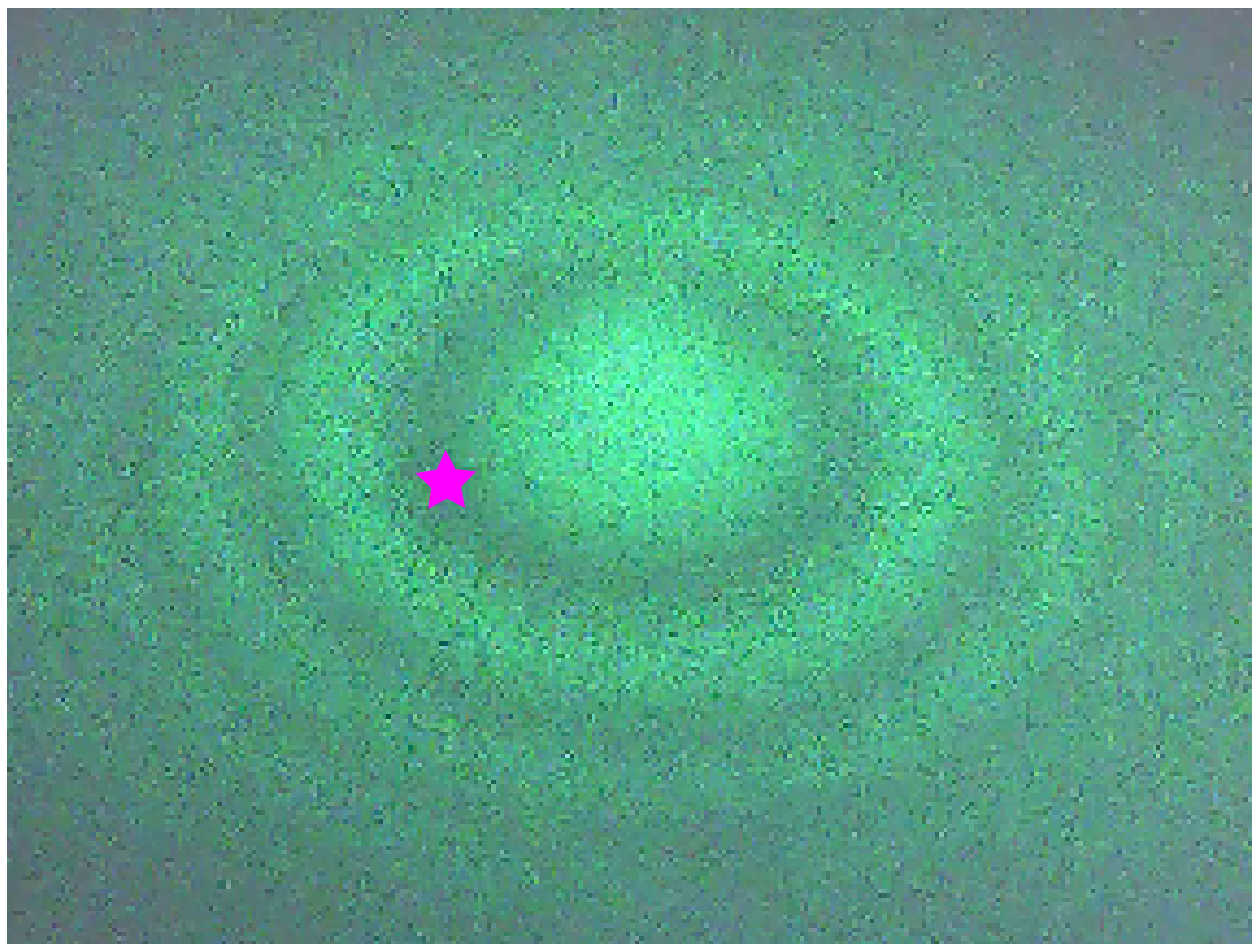

FIG. 2. The interference pattern produced by the tabletop Michelson interferometer. This image was taken with the webcam aligned off the beam axis. The pink star indicates the point where the intensity timeseries measurements were taken. The central bright fringe of the interference pattern is approximately 5 mm in diameter.

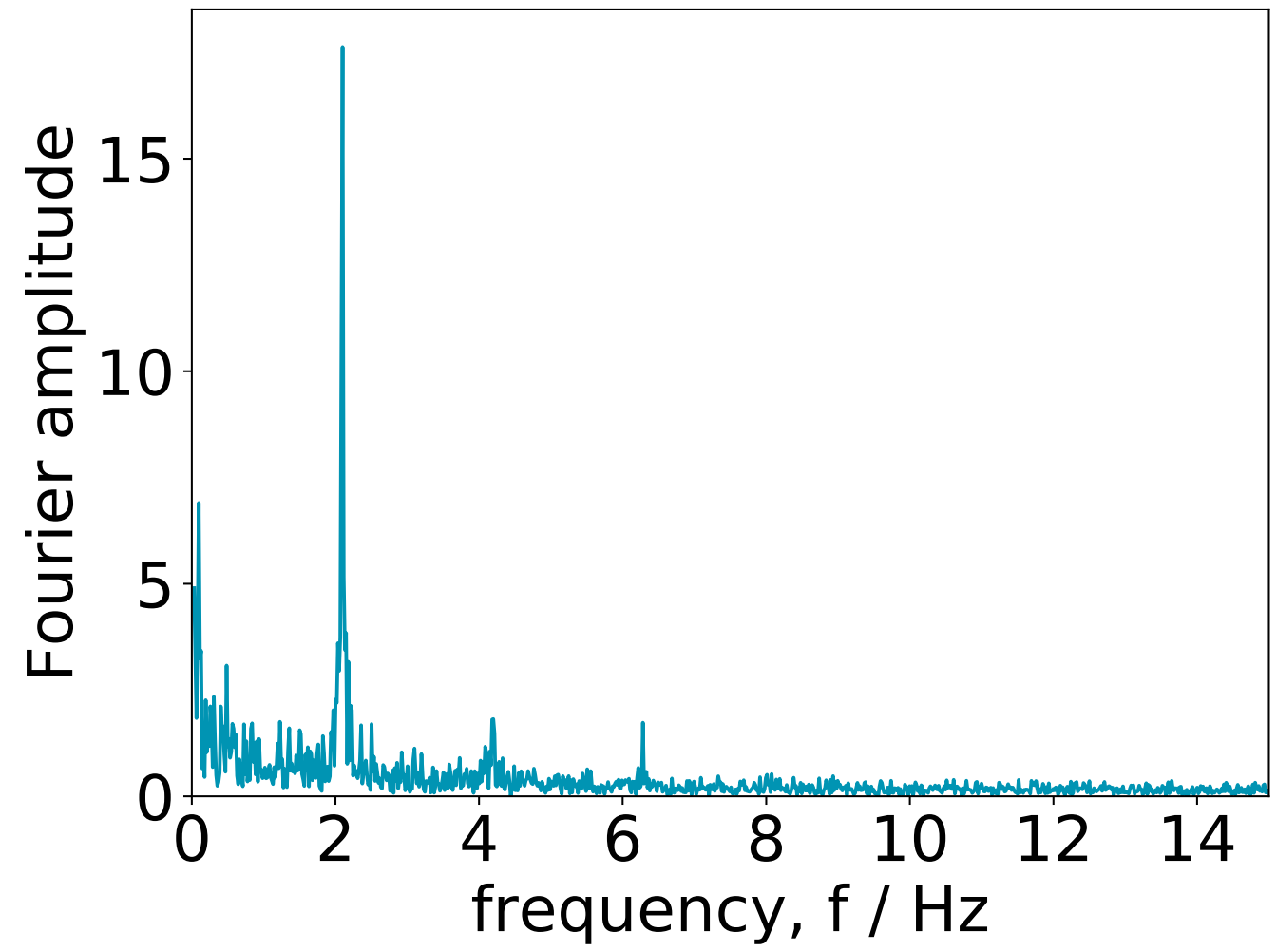

FIG. 3. Recovery of a tone at a constant frequency: the Fourier amplitude of the intensity pattern plotted against frequency. The injected signal has a frequency of 2.09 Hz, while the recovered signal peaks at 2.099 Hz and has an FWHM of 0.033 Hz. The plot also shows two harmonics at 4.19 Hz and 6.28 Hz with amplitudes 10.3% and 9.8% of the amplitude of the primary peak, respectively.

The motion of the interference fringes follows the motion of M2, and therefore of the speaker. The amplitude of these motions is given by a transfer function that accounts for the coupling and resonance of the speaker-mirror system. If the relative change in the length of the arms is kept small enough, then the intensity of the interference pattern at any point on the screen oscillates at the same frequency as the injected audio. For larger relative length changes, multiple fringes will pass through the detection point during a single speaker oscillation, raising the measured frequency artificially. As such, any motion of the fringes must be kept small by playing the sound softly through the speaker. Even without overcounting, large fringe motions display a nonlinear relation between the intensity fluctuations and injected audio, leading to troublesome—but physically interesting—phenomena like frequency doubling.

## III. CONSTANT FREQUENCY SIGNAL

Continuous-wave searches look for nearly monochromatic signals.[9] In this section, we consider a simple sinusoidal tone at a single, constant frequency.

As described in Section II, the audio signal is played through a speaker fixed to the back of M2 (see Fig. 1). The intensity of the interference pattern is measured at a single point on the screen, indicated by the pink star in Fig. 2. The webcam records video in three color channels: red, blue, and green. We use the green channel as an approximation of the total intensity produced by the green laser. The webcam samples at a rate of 30 Hz, which limits the spectral content of observable signals to less than 15 Hz, the Nyquist frequency.

A tone with a frequency of 2.09 Hz is played through the speaker for one minute. The amplitude of the Fourier transform of the recovered signal is shown in Fig. 3. The discrete Fourier transform is calculated using the NumPy package for Python (see Appendix A). We measured a peak amplitude at 2.099 Hz with a full width half maximum (FWHM) of 0.033 Hz. Two harmonics can also be seen at integer multiples of the peak frequency. The peaks at 4.19 Hz and 6.28 Hz have amplitudes of 0.103 and 0.098 as a fraction of the height of the main peak, respectively. They are likely due to the nonlinear response of the system discussed in Section II. Also, note that the noise appears to rise at lower frequencies; however, the origin of this noise is not known.

## IV. WANDERING FREQUENCY SIGNAL

A continuous gravitational-wave signal may wander slowly (and randomly) in frequency over time, due to stochastic internal processes in the superfluid interior of isolated neutron stars, or variable accretion from a stellar companion for neutron stars in binaries such as LMXBs (see Ref. [7] and references therein). Although continuous gravitational waves are nearly monochromatic, the long observing times ($\sim 1$ yr) mean that searches can be impacted by very small frequency drifts. A typical observation involves $1 \times 10^{10}$ wave cycles, and a coherent search must track the phase to better than half a cycle over the full observation. Here we consider the audio analog of a tone that wanders in frequency.

A Fourier transform applied to the whole dataset (as in Section III) is not well suited to the wandering frequency case as the signal is spread across multiple frequency bins. In continuous-wave searches, the wandering frequency problem is solved using the Viterbi algorithm,[10] which can track the signal's frequency over time. The analysis described and presented in this section is directly inspired by real continuous-wave search methods, [7] yet is pitched at a level appropriate for an undergraduate laboratory setting. In Section IV A, we briefly review the analysis method used by LIGO and Virgo; in Section IV B, we describe the methods as applied in this work; and in Section IV C, we show results for recovering a wandering frequency signal using the table-top interferometer.

### A. Continuous wave analysis with real data

The methods used here are inspired by LIGO and Virgo continuous-wave searches. For further details on these methods and continuous-wave searches, we refer the reader to the references in the Supplementary Material.

Continuous-wave searches are performed on long datasets, months to years in duration. The frequency of the signal can wander significantly over the observation period. In this context, "significantly" means across multiple frequency bins, where the typical width of a frequency bin is the reciprocal of the total observation time.[6, 9]

A "hidden Markov model" can be used to search for continuous gravitational waves.[7] In a Markov process, the current state depends only on the previous state (in this case the "state" is the frequency of the signal). In a hidden Markov model, the frequency state of the signal is unknown (hidden) and can undergo transitions at discrete times. The transitions are Markovian in that the hidden state (i.e. frequency) of the system at any time depends solely on its state at the previous time.

A detection statistic relates the observed data to the hidden state and quantifies the likelihood of a signal being present in the data at each frequency and time bin. This likelihood is also called the emission probability in gravitational-wave literature. In gravitational-wave data analysis, the detection statistic gives the likelihood of a signal given the antenna beam pattern of the detector, which varies as the Earth rotates and orbits the Sun.[9] When searching for continuous waves from a neutron star in a binary (such as an LMXB), the Doppler modulation of the source also needs to be taken into account and a different detection statistic is used.[7]

In continuous-wave searches, a physical model of the target informs how far the frequency of the signal can wander over time. This is called the transition probability matrix. For example, in LMXB searches, the transition matrix allows the signal frequency to (i) stay in the same frequency bin, (ii) move up a single frequency bin, or (iii) move down a frequency bin in each time step.[6] In supernova remnant searches, source frequency is expected to decrease over time, therefore the allowable transitions are to (i) remain in the same frequency bin, or (ii) move down one frequency bin (see the Supplementary Material for LMXB and supernova remnant search references).

The Viterbi algorithm [10] is used to find the most probable sequence of hidden frequency states given the sequence of observables. In the next section, we describe our application of the Viterbi method and our choice of detection statistic.

### B. The hidden Markov model and Viterbi algorithm

First, we split the timeseries data from the interferometer into segments. Then, we take the discrete Fourier transform of each timeseries segment to form a grid in time and frequency (a spectrogram) with the Fourier amplitude $F(t_i, f_j)$ at frequency $f_j$ and time $t_i$, as the detection statistic (this is the emission probability). Assuming Gaussian noise, this detection statistic maximizes the likelihood of detecting a sinusoidal signal as described in Appendix B. The detection statistic is normalized for convenience by dividing each value by the maximum Fourier amplitude in the grid (such that $\max_{i,j} F(t_i, f_j) = 1$).

Fig. 4 represents a spectrogram with $N_t$ time bins and $N_f$ frequency bins. The circular nodes represent the ele-ments of the time and frequency grid, which are labelled as $t_i$ and $f_j$, respectively, where $i = 0, 1, 2, \ldots, N_t$ and $j = 0, 1, 2, \ldots, N_f$. The shading of the nodes represents the detection statistic $F(t_i, f_j)$ (the observable) at each grid point, where dark corresponds to lower values and light to higher values.

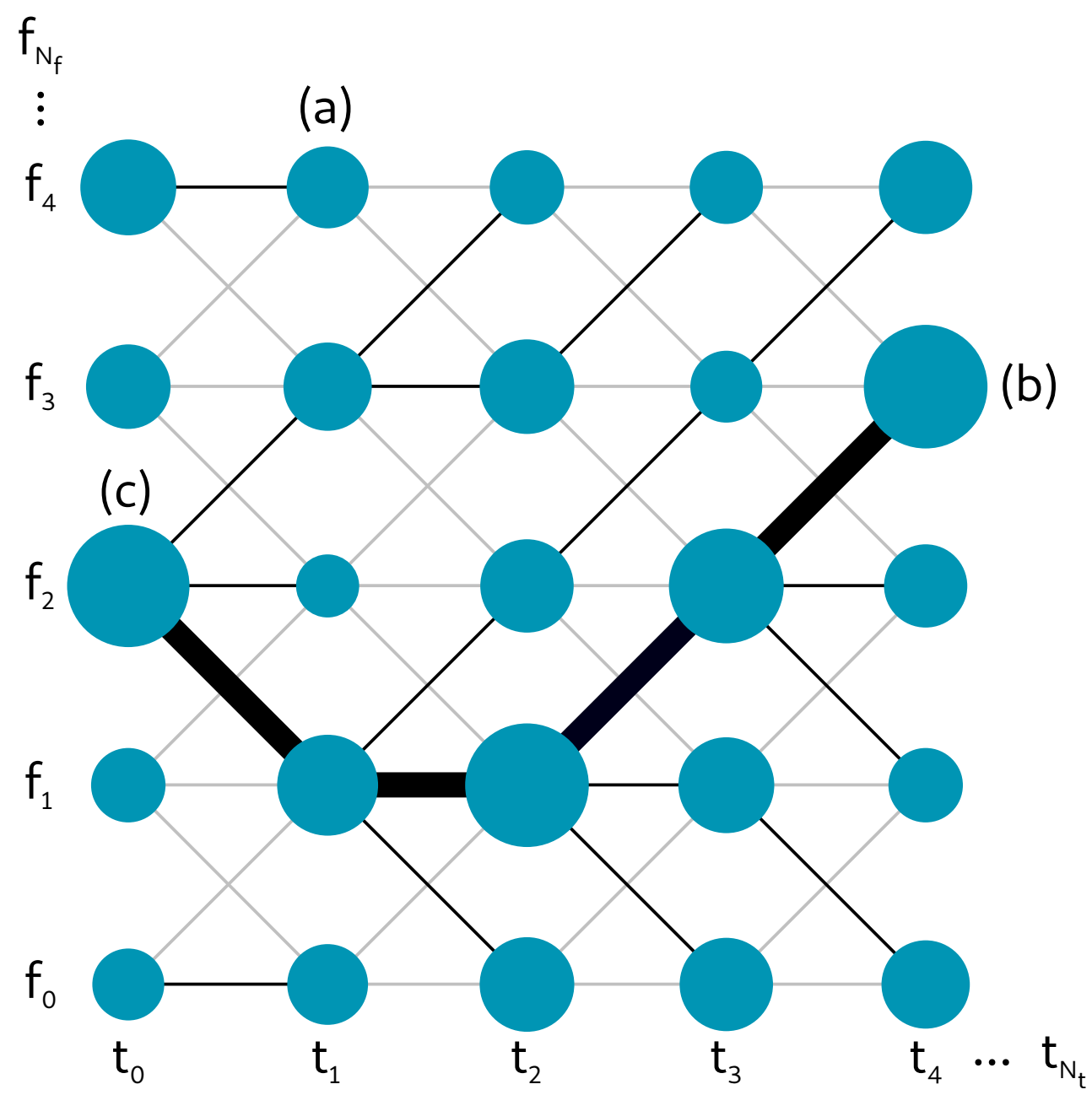

FIG. 4. Schematic diagram of the Viterbi algorithm. The circular nodes represent elements in a time-frequency grid, labeled $t_0$ to $t_{N_t}$ (left to right) in time and $f_0$ to $f_{N_f}$ (bottom to top) in frequency. The size of the nodes represents the likelihood of a signal being present in each time-frequency element. The small to large sizing corresponds to low to high likelihood values in arbitrary units for this schematic. The objective is to find the most probable path of a signal through the grid from left to right. All possible paths are shown by the lines. The black lines show the best path to each node and the gray lines show rejected paths. Some routes through the grid result in dead-ends with respect to optimally reaching the other side, such as the path ending at $t_1$ marked (a). At $t = t_{N_t}$ the algorithm chooses the terminating frequency node which has the highest value given by Eq. 1, marked (b). The Viterbi path is the path leading to this node, highlighted by the thick-black lines, from (b) to the start at (c).

The objective is to find the most likely path through the grid, given the observed data and any probabilistic constraints on how the frequency of the signal can wander from $t_i$ to $t_{i+1}$. The transition probability matrix, $A(f_k, f_m)$, describes the probability of the system transitioning from state $f_k$ at $t_i$ to a state $f_m$ at $t_{i+1}$. Here, we allow the frequency of the signal to either (i) stay in the same bin, (ii) move up by a single frequency bin, or (iii) move down by a single frequency bin at each transition. We assign these three transitions equal probability, i.e. $A(f_k, f_m) = 1/3$ for $k = m + 1, m, m - 1$ and $A(f_k, f_m) = 0$ otherwise. All possible transitions are shown as lines in Fig. 4. The different line colors are explained below.

Before we begin analyzing the data, we have no prior knowledge as to which frequency bin the signal starts in (i.e. the prior is flat between the minimum and maximum frequency bins). At the first stage of the analysis, we define the probability of the system having frequency $f_j$ at the initial time $t_0$ to be equal to the (normalized) detection statistic of that state (i.e. $\Pr[f(t_0) = f_j] = F[t_0, f_j]$). A specific path (which may not be optimal) is written as $f(t_0), f(t_1), \ldots, f(t_n)$. The probability of a specific path given the data is

$$\begin{aligned} \Pr[f(t_0), f(t_1), \ldots, f(t_n)|\text{data}] & \\ = F[t_n, f(t_n)] A[f(t_n), f(t_{n-1})] & \\ \times \ldots \times A[f(t_1), f(t_0)] F[t_0, f(t_0)] \,. & \quad (1) \end{aligned}$$

The path $f^*(t_0), \ldots, f^*(t_n)$ that maximises $\Pr[f(t_0), f(t_1), \ldots, f(t_n)|\text{data}]$ is the optimal path terminating in the frequency bin $f^*(t_n)$. We note that the left-hand-side and right-hand-side of Eq. 1 are both evaluated for a specific path and then we maximize over all such paths to find the optimal Viterbi path.

The Viterbi algorithm provides a computationally efficient method for finding the optimal path. At every $t_i$ all but $N_f$ possible paths are eliminated. Here we describe the algorithm while referring to the schematic in Fig. 4. The implementation used in this work is available online (see Appendix A) and we provide further information (including pseudocode) in Appendix C.

1. Starting at time $t_1$, each $f_j$ node (circles in Fig 4) can originate from three prior nodes at time $t_0$ (except for the edge cases $f_0$ and $f_{N_f}$ which only have two). The paths between these nodes are indicated by the lines in Fig. 4. At each $f_j$ node, we select the path with the highest $A[f(t_0), f(t_1)]\Pr[f(t_0)]$ value as the most probable path. These choices are highlighted using the black lines in Fig. 4 while the gray lines show the rejected paths. For example, the most probable connection to the node labeled (a) is the one directly behind it (i.e., $f_4$). Therefore, this path is selected as the best path from $t_0$ to $t_1$ for $f_4$. To allow backtracking at the end, the index of the most probable connection to a node along with the value of the best path to that node is stored, for each node.

2. Moving to time $t_2$, again we select the path which maximizes Eq. 1 for each $f_j$. These paths are again shown by the solid black lines between the nodes at $t_1$ and $t_2$ in Fig. 4. Rejected paths are again shown by gray lines.

3. Step 2 is repeated until the end of the grid ($t = t_{N_t}$) is reached with only the best paths being stored at each iteration.

4. We have now found the most probable path to each $f_j$ at $t = t_{N_t}$ and its probability (Eq. 1). We se-

lect the terminating frequency bin $f(t_{N_t})$ with the highest probability labeled as (b) in Fig. 4.

5. The final step is to find the Viterbi path (the overall best path that terminates in the frequency bin with the highest probability in step 4). The Viterbi path is found by backtracking along the stored best connections at each $t_i$ (see also Appendix C). In Fig. 4, it is the path ending at (b) that started at (c) highlighted by the thick-black lines.

In continuous gravitational wave searches, the signal amplitude is expected to be small in comparison to the noise and its frequency can change unpredictably over time. The Viterbi algorithm's strength lies in its ability to track such signals through the data even in the presence of comparatively loud noise fluctuations in the time-frequency bins. In the following section, we present the results of using the Viterbi algorithm with the table-top interferometer data.

### C. Wandering frequency signal results

We simulate a slowly wandering signal by modulating the frequency sinusoidally with a modulation amplitude that decays with time. We use the same apparatus as shown in Fig. 1 and the output interference pattern is recorded via webcam as in Section III. In this section, we test the Viterbi algorithm's ability to recover the wandering signal. Note that we implicitly approximate the noise at the webcam as white (uniform in frequency) in this implementation of the Viterbi algorithm. This approximation ignores the increase in noise at low frequencies (below 1 Hz) shown in Fig. 3. However, white noise is a reasonable approximation for the narrow frequency range (around 3–11 Hz) we use in this analysis.

The results are shown in Fig. 5, where the heat-map shows the spectrogram of the observed signal (similar to that represented by the schematic in Fig. 4). In this demonstration we use a signal that can easily be identified in the spectrogram; however, we expect real continuous-wave signals to have far lower signal-to-noise ratios at each time step. The overlaid pink dots in Fig. 5 show the injected signal and the white crosses show the recovered Viterbi path. The recovered Viterbi path is within one frequency bin ($\sim 0.3\,\mathrm{Hz}$) of the injected signal for 94% of the time. We also compute the root-mean-square (RMS) fractional error $E_{\mathrm{rms}}$ along the path, which is defined as

$$E_{\mathrm{rms}} = \left[\frac{1}{N_t+1}\sum_{i=0}^{N_t}\frac{(I_i - R_i)^2}{{I_i}^2}\right]^{1/2}, \qquad (2)$$

where $I_i$ and $R_i$ are the injected and recovered (Viterbi) frequencies at time $t_i$, respectively. For the result shown in Fig. 5, we find that $E_{\mathrm{rms}} = 0.082$, which indicates fair but not total recovery of the injected signal.

This may be explained by two anomalies in the recovered path. Initially, the injected signal wanders by more than one frequency bin per time bin (i.e. faster than our choice of $A(f_k, f_m)$ allows), thus leading to a discrepancy between the injected and recovered paths for $t \lesssim 10\,\mathrm{s}$. If we remove the first four time bins from the $E_{\mathrm{rms}}$ calculation, we find a slight improvement in recovery with $E_{\mathrm{rms}} = 0.044$. One may be tempted to increase the allowed frequency wander in the algorithm; however, this leads to an overall decrease in the above statistics, as the algorithm is prone to jump briefly to nearby spots of noise. There is also an anomaly at 150 s, which is likely due to a local disturbance, e.g. someone walking past the interferometer. As shown in Fig. 5, the Viterbi algorithm can recover and continue to track the signal after the disturbance.

## V. COMPLEX AUDIO: MUSIC AND SPEECH

In this section, we explore how our apparatus can be used to teach a selection of signal processing techniques. We use complex audio signals (such as music and speech) as natural successors to the constant and wandering tones used in Sections III and IV, respectively. As complex audio signals are not quasi-monochromatic, the Viterbi algorithm used in Section IV is not directly applicable here. Instead, we use a hierarchy of passive filters which suppress noise, yet do not assume any specific form of the signal, unlike the Fourier-based maximum likelihood filter which is tuned to the sinusoidal signals in Section III and Appendix B.

We use the Michelson interferometer as an "optical microphone" to detect sound, replacing the components of a conventional microphone with a laser interferometer. The only change to our apparatus is replacing the webcam with a photodiode to allow us to capture the higher frequencies necessary for speech and music (see Section V A). Optical microphones have precedence in the laser microphones [11] which are (or were historically) used in the defense industry and operate on a variety of related principles. Our objective is to play a recording of speech or music through the speaker attached to mirror M2 (see Fig. 1), record the resulting interference pattern, and then recover the original signal via a selection of signal processing techniques.

The apparatus serves as an independent demonstration for a broader physics and engineering audience, particularly in undergraduate laboratories. We describe additional hardware components required for this demonstration in Section V A and the initial results in Section V B. We consider a selection of filter techniques, details of which, along with a summary of digital signal processing resources, can be found in the Supplementary Material. In Section V C, we present the two best-performing techniques from the Supplementary Material.

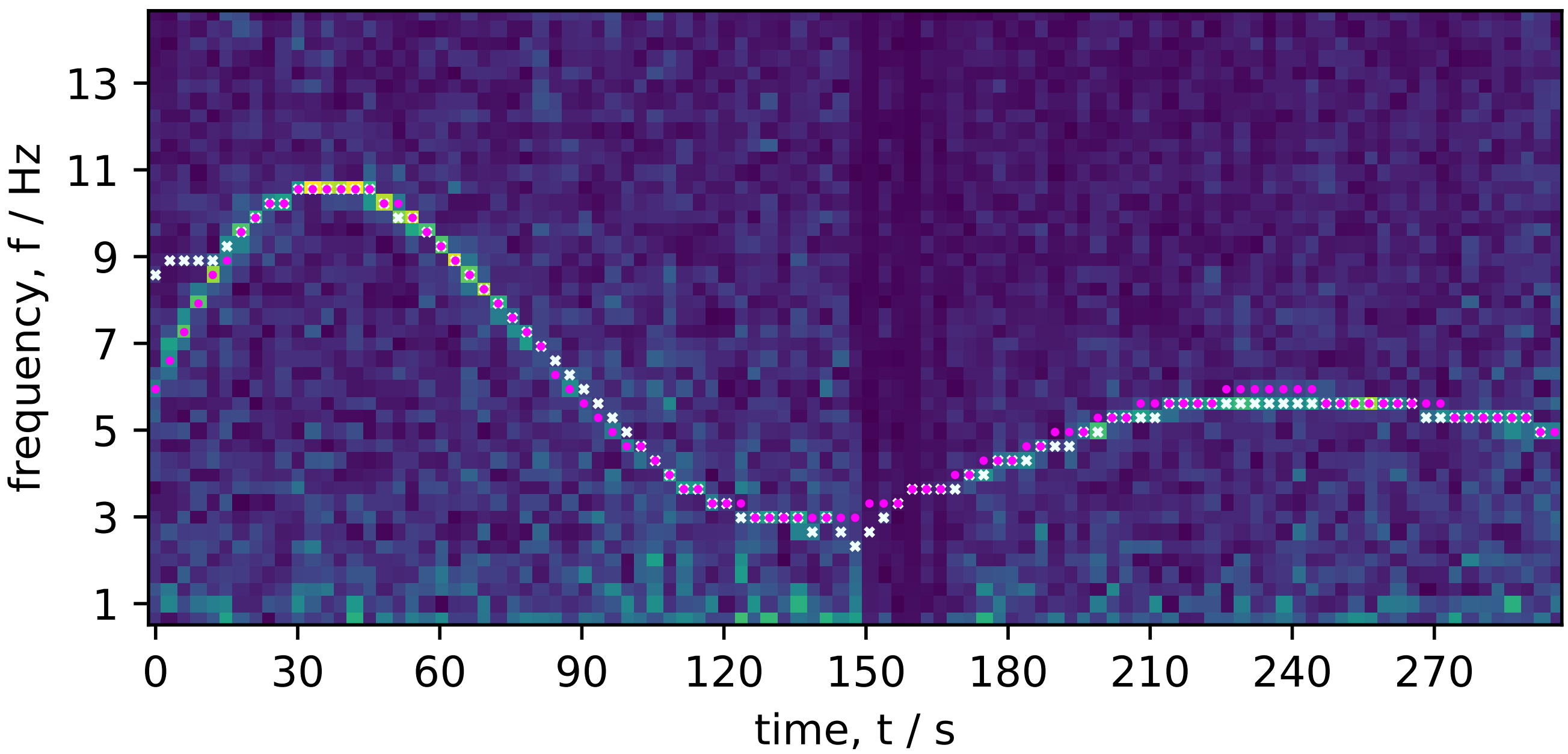

FIG. 5. Recovery of a wandering tone using the Viterbi algorithm. The spectrogram (heat-map) coloring indicates the value of the detection statistic (here, the logarithm of the absolute value of the discrete Fourier transform) in each time-frequency bin – with brighter colors indicating higher values. The overlaid pink-dot and white-cross markers show the injected signal and recovered Viterbi path, respectively. On the left, before $\sim 15\,$s, the signal changes frequency too quickly for the Viterbi algorithm to recover, given that the algorithm is restricted to only change by one frequency bin per time bin. At 150 s, the data appears anomalous, which may be due to some transient background noise.

### A. Hardware modifications for the optical microphone

The human ear can hear frequencies in the range of $\sim 20\,$Hz–20 kHz. Speech intelligibility (the ability to understand speech) requires frequencies up to 3 kHz and music requires up to and beyond 8 kHz. Therefore, the optical microphone requires a sample rate of at least 16 kHz to capture both speech and music (adjusting for the Nyquist frequency). This cannot be achieved with the webcam used in Sections III and IV as it has a sampling rate of 30 Hz and thus can only "hear" frequencies below 15 Hz. To overcome this issue, we use a photodiode [12] at the output of the interferometer to achieve a sampling rate of 16 kHz.

We place an OSRAM BPW21 photodiode in reverse-bias over an LM358 op-amp which together produce a voltage that depends on the incident intensity. The photodiode records the interference pattern at roughly the same off-center position as the webcam in Sections III and IV, again chosen arbitrarily. The voltage signal from the photo-detector is captured by an MCP3008 10-bit analog-to-digital converter (ADC) connected to a Raspberry Pi Model 3 v1.2, which provides a convenient means to record the photodiode data. Together, the circuit samples the signal at $\sim 16\,$kHz. Resources for using the Raspberry Pi and photodiode, including a circuit diagram, are described in the Supplementary Material.

Sampling any frequency component of the analog signal above the Nyquist frequency of 8 kHz leads to aliasing (folding of frequencies greater than half the sampling rate) into the detected range. We include an anti-aliasing Sallen-Key filter with a cut-off frequency of 8 kHz before the ADC to prevent this from happening. This component attenuates any frequencies above 8 kHz before they are digitally sampled. We also place a cloth screen over the face of the photodiode to reduce the incident intensity and avoid saturating the ADC – an improvised, physical solution that could instead be replaced by scaling down the voltage electronically. This cloth screen was re-purposed grill cloth from a commercial speaker.

### B. Anti-aliased output

We test the optical microphone with a variety of recordings, including the speech of different people and music ranging from simple melodies and rhythms to songs. During recordings, care is taken to minimize activity around the demonstration to reduce environmental noise coupling into the interferometer. The timeseries data is then directly converted to a .wav file and played as an audio recording using the `scipy.io.wavfile.write` function in Python (see Appendix A). When processing the results, we restrict our analysis to only the first 10 s of each observation (for efficiency), and only plot the first second in our results (Fig. 7).

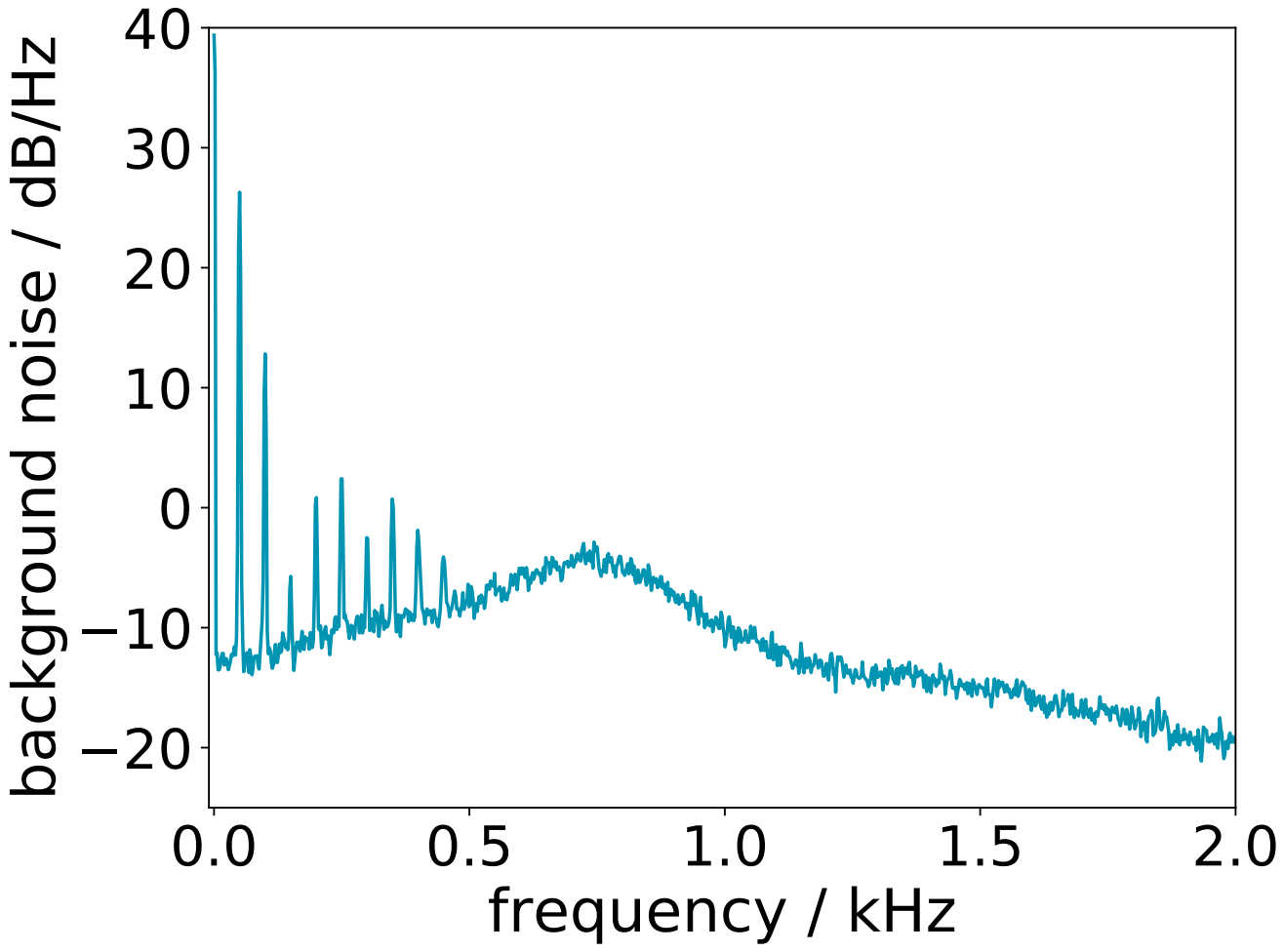

FIG. 6. Power spectral density (PSD) of background noise from the optical microphone (with the speaker off). We see strong power from the 50 Hz mains hum and its harmonics (most likely from the photodiode circuit and the room's lighting and cooling). Otherwise, the PSD is fairly white except for a peak at around 0.75 kHz.

The raw output of the optical microphone (with anti-aliasing) is noisy with a loud, continuous bass hum. This can be explained by looking at the power spectral density (PSD) of the background noise (i.e., the output with the speaker switched off), shown in Fig. 6. The spectrum is dominated by AC eletrical power grid noise from the fundamental 50 Hz Australian mains electricity grid signal up to and beyond the 8th harmonic (at 400 Hz). The mains signal is also present, but far weaker, in the background spectrum taken with the photodiode in darkness, suggesting that ambient lighting has a large contribution. Besides lighting, other possible contributions to the mains signal include air conditioning and the photodiode circuit itself. The appearance of harmonics of the mains noise might be due to the non-linearity in the system discussed in Section II. The spectrum in Fig. 6 also has a broad feature at around 750 Hz, the origin of which is yet to be determined. Environmental noise reduction for gravitational-wave detectors is an active area of research (see the Supplementary Material for further information and resources on this topic).

### C. Optical microphone results

We explore several filters to remove the 50 Hz mains hum and harmonics and improve the speech intelligibility of the recording. The Supplementary Material describes a range of analysis techniques that can be used as examples for the undergraduate laboratory. All filters are tested on the same 1 s long speech recording. The results of this section are shown in Fig. 7. In the figure, the left and right columns show the timeseries and frequency spectrum, respectively. The first row shows the input signal played through the speaker (see Fig. 1). The second row shows the raw output from the photodiode recording.

In signal processing, the ideal filter would be one that: (i) completely attenuates the undesired parts of the spectrum, (ii) does not change the rest of the spectrum, and (iii) smoothly transitions between these regions, as to not damage the time domain signal when seen under convolution. However, these three conditions cannot all hold at once. For example, if conditions (i) and (ii) hold, then the filter must be discontinuous at the boundary of the undesired region but this implies that the filter has "infinite latency" and so will affect (or damage) the time domain signal for infinite time.[13] Therefore, any filter must compromise between these three conditions. For speech intelligibility, this means that either: (i) some noise remains in the filtered recording, (ii) some of the speech content is lost as certain important frequencies are attenuated, or (iii) the speech is somewhat distorted in time. All three of these cases can, when taken to the extreme, make the speech intelligibility worse than the unfiltered recording. Therefore, we choose filters that compromise between achieving the three conditions.

In this section, we present the results of two advanced signal processing techniques applied to the optical microphone recordings. The techniques are only briefly described here and we refer the reader to the Supplementary Material for further details and other analysis techniques.

Firstly, we consider two signal processing techniques used in combination: the cascaded notch and the Wiener filter (see also the Supplementary Material). A notch filter removes signals within a specific frequency range. We want to remove the 50 Hz mains noise and harmonics, therefore we use a cascaded notch filter where each notch is centered on one of the harmonics. The Wiener filter is an advanced statistical technique that makes use of statistical information from the speech data and noise. It amplifies parts of the signal with a high signal-to-noise ratio while suppressing parts with a low signal-to-noise ratio. The results of the combined cascaded notch and Wiener filter are shown in the third row in Fig. 7. Most of the mains noise is removed; however, the recovered voice sounds muffled and is not understandable.

Secondly, we apply a speech enhancement technique. Ref. [14] compares 13 speech enhancement methods, finding the log minimum mean-square error (logMMSE) estimator to be the best, qualitatively, at recovering speech (see also the Supplementary Material). We use an existing implementation of the logMMSE from Ref. [15]. The logMMSE estimator results are shown in the bottom panels in Fig. 7. We see significant attenuation of the mains harmonics and general smoothing of the spectrum. Most of the background noise is removed; however, the logMMSE still does not significantly enhance the speech as the voice sounds muffled and indistinct.

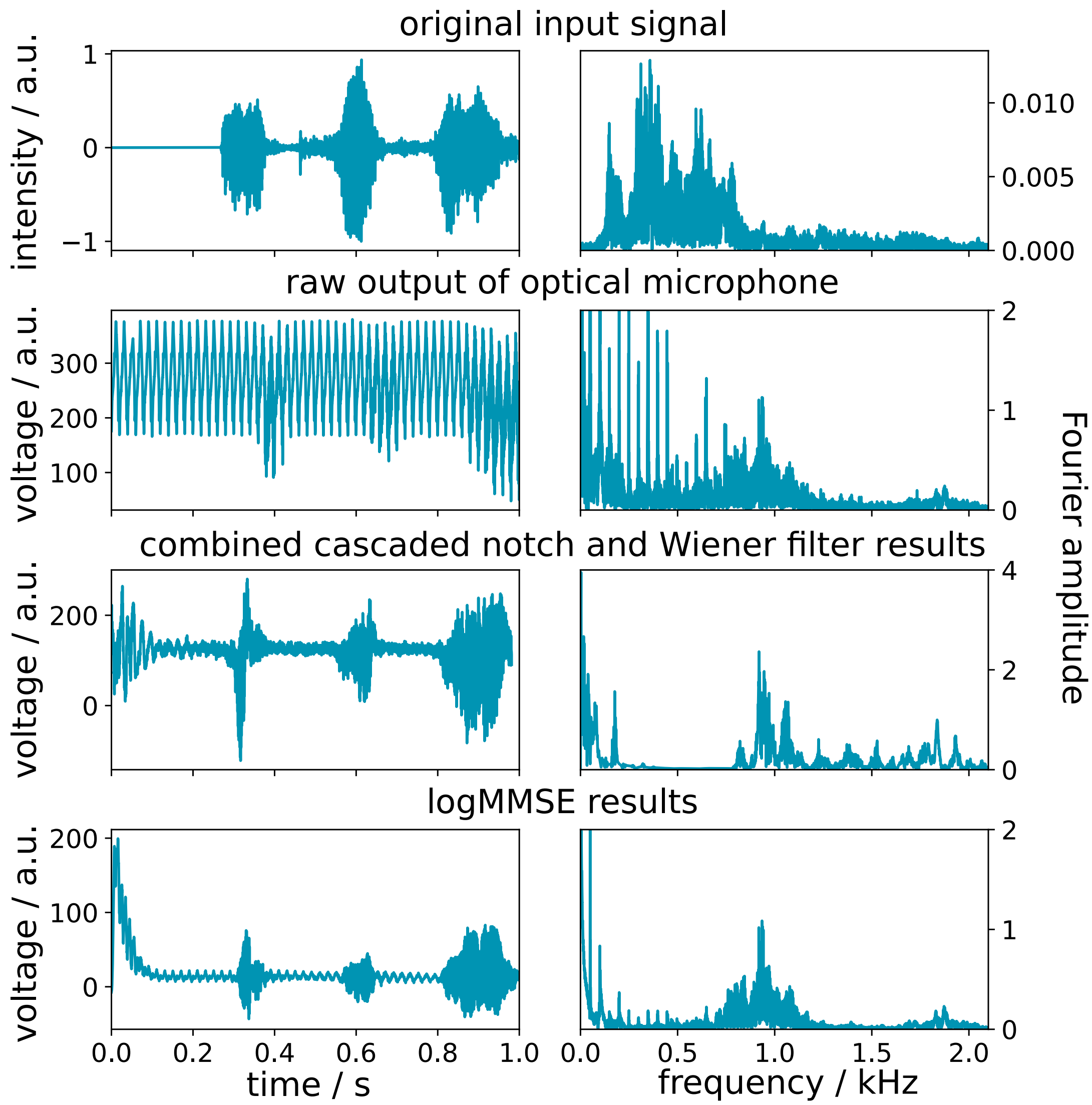

FIG. 7. Timeseries (left column) and frequency spectrum (right column) results with the optical microphone. The original input signal in the first row is a 1 s recording of an adult male voice (saying "a cathode"). The input signal is shifted by 0.12 s to the right to synchronize the manual delay from starting the recording with the Raspberry Pi and starting to play the source through the speaker. The second row shows the raw output from the optical microphone when the input from the first row is played. The third row shows the result of applying the notch and Wiener filters combined. The fourth row shows the result of applying the logMMSE estimator, where the rise at the start of the timeseries is an expected effect when filtering a signal of finite duration.

We find some improvement with music over speech. Simple chords and drums can be heard after filtering, but more composite sounds and complex melodies cannot be heard clearly. Our observations suggest that this is especially true for certain instruments; in particular flutes and violins sometimes can't be heard at all. This could be a perceptual effect or a frequency dependence somewhere in the optical microphone. Speculating, perhaps the speaker-mirror coupling is stronger at low frequencies and thus instruments like electric bass and drums sound louder in the results. To address these problems, we need to determine whether the signals that are audibly missing (the diction in the speech and complex melodies in music) are indeed being transmitted through the optical microphone at all. To determine this requires a better understanding of the system, as discussed in Section VI.

## VI. FUTURE WORK

The experiment described here can be used to demonstrate and teach a variety of topics, from basic optics and photodiode circuits, to signal processing and speech en-hancement. Cross-disciplinary undergraduate laboratory experiments between physics and engineering courses are beneficial to student learning (see Section II in the Supplementary Material for further information). Here, we consider further extensions and improvements to the experiment such as noise isolation, transfer function analysis, speech recognition, and control systems.

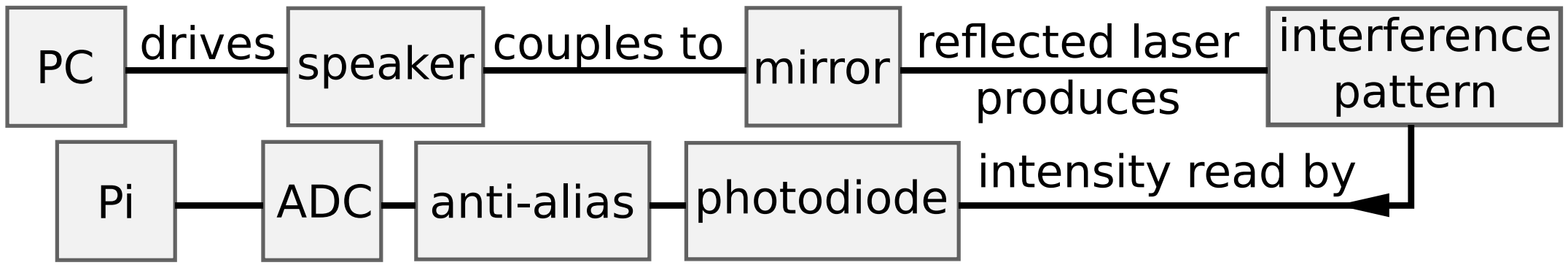

FIG. 8. Flowchart for the signal through the optical microphone. A transfer function of the system would have to account for each stage and exploration of this is an aim of future work. Not shown are the signal processing filters that are performed on the PC once the data is transferred off the Raspberry Pi.

Isolating the experiment from interference may provide some noise reduction. The Raspberry Pi could be powered with a commercial battery instead of mains power, which may help to isolate the circuit from the mains as all other components draw power from the Raspberry Pi (the laser is DC-powered through a converter). Transferring the circuit from a breadboard to a printed circuit board can also reduce interference.[16] A feedback loop control system can also be used to suppress any unwanted movement of the interferometer components,[17–19] similar to methods used to isolate gravitational-wave observatories.

A thorough characterization of the system's transfer function would allow us to understand how audio signals couple through the interferometer. The total transfer function starts with the voltage being sent to the speaker and ends with the voltage recorded by the Raspberry Pi. It would include any inherent non-linearities in the speaker, the speaker-mirror coupling, the path length to intensity relation, and the photodiode. The schematic in Fig. 8 shows the abstract signal flow through the system. It does not include other potential pathways for signal flow such as acoustomechanical vibrations of the webcam or photodiode mount which could also be examined.

A voluminous literature exists on using hidden Markov models trained on phonemes to recognize speech.[20] These could perform the final stage of speech enhancement for the optical microphone. This coincidentally connects back to the Viterbi algorithm in Section IV, which is also underpinned by a hidden Markov model, albeit of a different type. Alternatively, machine learning solutions exist throughout the field that can provide results competitive with statistical techniques (such as the log-MMSE estimator used here).[21]

More traditional techniques such as a wavelet transform [22] could be used to extract the signal from noise and compared with the above methods. A wavelet transform provides both time and frequency information, making it easier to pinpoint the origin of noise with respect to time. In Refs. [23, 24], wavelet transform methods are proposed for speech recognition.

Returning to the demonstration of gravitational wave analysis, single interferometers cannot yield directional information for signals; a large proportion of the directional information in gravitational-wave detections comes from the time delay offset between the observation being recorded at two or more detectors.[1] We could therefore extend this analysis to include data from two interferometers to extract directional information. This would require increasing the sensitivity of the optical microphone to pick up the signal from a distant source instead of from a speaker attached directly to one of the mirrors of the interferometer. Another extension would be to demonstrate the Doppler effect of the Earth's motion around the Sun, which needs to be considered in continuous-wave searches (see Section IV A and Ref. [9]). One approach could be to modify the input audio signal to simulate Doppler modulation.

## VII. CONCLUSIONS

In this paper, we use a table-top Michelson interferometer as an analog to a gravitational-wave detector, demonstrating signal processing techniques used within the gravitational-wave community. We explore the use of the interferometer as an optical microphone and consider a more general treatment of signal processing with complex audio signals, which can also serve as a distant analog to minimally modelled gravitational-wave burst signals, e.g., from supernovae. The demonstration can be adapted for use in both the physics and engineering undergraduate laboratory, providing opportunities for cross-disciplinary teaching. Additionally, it can be used as a tool for explaining gravitational-wave research to a wider, non-specialist audience.

As the field of gravitational-wave astrophysics continues to grow, the future will bring many more detections of binary black holes and neutron stars, as well as the anticipated first detection of other classes of signals, such as continuous waves, to which this demonstration provides some charming insights.

## ACKNOWLEDGMENTS

The authors are grateful to Jude Prezens, Alex Tolotchkoc, and Blake Molyneux for their technical advice and generous assistance; Patrick Meyers, Margaret Millhouse, Sofia Suvorova for useful discussions; Patrick Clearwater, Patrick Meyers, Suk Yee Yong, Lucy Strang, Julian Carlin, Sanjaykumar Patil, and Alex Cameron for early work on the interferometer design requirements and construction; and the LIGO Education Public Outreach working group, in particular, Anna Green, Lynn Cominsky, Sam Cooper, and Martin Hendry, for their helpful feedback and suggestions. This research is supported by the Australian Research Council Centre of Excellence for Gravitational Wave Discovery (OzGrav) (project number CE170100004). Financial support towards hardware was provided by the Institute of Physics International Member Grant and the OzGrav Outreach Support scheme. Travel support was provided by the Australian National University PhB Science program. This work has been assigned LIGO document number P2000386.

## Appendix A: Open-source code

This project is implemented in Python 3 scripts and jupyter notebooks and MATLAB. We refer the reader to the Supplementary Material for software references. The current build and sample data can be found at: `https://github.com/daccordeon/gravexplain`

## Appendix B: Detecting a sinusoidal signal in Gaussian noise

In this appendix, we demonstrate that the modulus of the Fourier transform is an appropriate detection statistic when searching for a sinusoidal signal in Gaussian noise. We describe the data as

$$x(t) = s(t) + n(t)\,, \tag{B1}$$

where $s(t)$ and $n(t)$ are the signal and noise, respectively. The signal takes the form

$$s(t) = A\cos\left[\omega t + \phi\right]\,, \tag{B2}$$

where $A$, $\omega$, and $\phi$ are the amplitude, angular frequency and phase of the signal, respectively. The noise $n(t)$ is a fluctuating zero-mean time series with the following property: if we define an inner product between two arbitrary time series $u(t)$ and $v(t)$ as

$$\langle u, v\rangle = \frac{1}{T}\int_0^T \mathrm{d}t\, u(t)v(t)\,, \tag{B3}$$

where $T$ is the total time of the observation, then the probability $\mathcal{L}$ of measuring the noise-noise product $\langle n, n\rangle$ is given by

$$\mathcal{L} = \exp\left(-\frac{1}{2}\langle n, n\rangle\right)\,. \tag{B4}$$

Equations B3 and B4 define what it means for noise to be Gaussian through the fundamental measurement of $\langle n, n\rangle$.

The likelihood of measuring the signal $s(t)$ in the presence of noise follows from Eqs. B1 and B4 by replacing $n(t)$ in Eq. B4 with $x(t)-s(t) = n(t)$ from Eq. B1. [9, 25] The result is

$$\mathcal{L} = \exp\left(-\frac{1}{2}\langle x-s, x-s\rangle\right)\,, \tag{B5}$$

$$= \exp\left(-\frac{1}{2}\langle x, x\rangle - \frac{A^2}{2}\langle\cos[\omega t+\phi], \cos[\omega t+\phi]\rangle + A\langle x, \cos[\omega t+\phi]\rangle\right)\,, \tag{B6}$$

$$= \exp\left(-\frac{1}{2}\langle x, x\rangle - \frac{A^2}{4} + A\langle x, \cos[\omega t+\phi]\rangle\right)\,. \tag{B7}$$

We then maximise Eq. B7 with respect to $A$, obtaining

$$\mathcal{L}_{\max} = \exp\left(-\frac{1}{2}\langle x, x\rangle + \langle x, \cos[\omega t+\phi]\rangle^2\right)\,, \tag{B8}$$

for $A = 2\langle x, \cos[\omega t+\phi]\rangle$. From the second term of Eq. B8, we see that the maximum likelihood of a sinusoidal signal in Gaussian noise is the modulus of the cosine Fourier transform, plus the term $\langle x, x\rangle$, which is independent of $\omega$ and $\phi$ and can therefore be ignored when searching over $\omega$.

Two important points must be made about the above procedure. (i) Fundamentally the goal is to maximize $\mathcal{L}$ in Eq. B5 by varying $s(t)$ through $A$. For the special case of the Gaussian likelihood (Eq. B4), which peaks at $\langle n, n\rangle = 0$, this is equivalent to minimizing the difference between $x(t)$ and $s(t)$ as evident in Eq. B3. In general, however, minimizing the difference between $x(t)$ and $s(t)$ is not equivalent always to the fundamental goal of maximising $\mathcal{L}$, for example if $\mathcal{L}$ peaks at $\langle n, n\rangle \neq 0$, or if $\mathcal{L}$ has multiple maxima. (ii) The maximum likelihood $\mathcal{L}_{\max}$ in Eq. B8 (or equivalently its logarithm) defines the detection statistic. When its value exceeds a threshold (chosen freely by the analyst) at some value of $\omega$, a signal is deemed to have been detected at that value of $\omega$. Therefore the specific functional form of Eq. B8 matters, which is a second reason why one must start from Eq. B4 rather than Eq. B5, in addition to reason (i).

## Appendix C: Viterbi algorithm

This appendix contains some details regarding the implementation of the Viterbi algorithm described in Section IV B. The Viterbi algorithm [10] is a classic method in signal processing, whose theoretical underpinnings and implementation are accessible to undergraduate students.

See the Supplementary Material for further resources and other pseudocode examples available online.

Here, we present some pseudocode (below) of the implementation used in Section IV B. We use Fourier amplitudes, normalized between $(0, 1)$ by dividing by the maximum value in the grid, as multiplicative weights. To avoid numerical underflow we take the logarithm of the Fourier amplitudes, which we can equivalently use as additive weights.

Let $X$ be a grid of $j = 0, \ldots, N_f$ rows and $i = 0, \ldots, N_t$ columns of additive weights for each node. Let Y and Z be grids of the same shape to store the weight of the best path to each node and the row index of the previous node on that path, respectively. Let W be a length $N_t+1$ array to store the final sequence of row indices for the optimal path. We restrict paths to only move one cell up or down at a time (or stay constant). For the boundary cases of $j = 0, N_f$, we only search over $k \in \{0, 1\}$ and $k \in \{-1, 0\}$, respectively, to stay inside the grid (this is not shown in the pseudocode below).

**function** VITERBI($X$)
  **for** each row $j = 0, \ldots, N_f$ **do**
    $Y_{0,j} \leftarrow X_{0,j}$
  **end for**
  **for** each column $i = 1, \ldots, N_t$ **do**
    **for** each row $j = 0, \ldots, N_f$ **do**
      $Y_{i,j} \leftarrow X_{i,j} + \max_{k \in \{-1,0,1\}} (Y_{i-1,j+k})$
      $Z_{i,j} \leftarrow j + \operatorname{arg\,max}_{k \in \{-1,0,1\}} (Y_{i-1,j+k})$
    **end for**
  **end for**
  $W_{N_t} \leftarrow \operatorname{arg\,max}_{j=0,\ldots,N_f} (Y_{N_t,j})$
  **for** each col $i = N_t - 1, \ldots, 0$ **do**
    $W_i \leftarrow Z_{i+1,W_{i+1}}$
  **end for**
  **return** $W$
**end function**

# Supplementary Material for
# Continuous gravitational waves in the lab: recovering audio signals with a table-top optical microphone

James W. Gardner, Hannah Middleton, Changrong Liu, Andrew Melatos, Robin Evans, William Moran, Deeksha Beniwal, Huy Tuong Cao, Craig Ingram, Daniel Brown, Sebastian Ng

## S1 Further reading and resources

Here we provide a selection of resources and reading material on topics related to this work.

### S1.1 Gravitational-wave observatories

The current network of ground-based gravitational-wave observatories includes the Advanced Laser Interferometer Gravitational-wave Observatory (LIGO [S1] which includes LIGO-Hanford and LIGO-Livingston in the United States and LIGO India which is under construction in India [S2]), Advanced Virgo [S3], GEO600 [S4], and KAGRA [S5].

### S1.2 Gravitational-wave observations

Gravitational waves have been observed from compact binary coalescences [S6]. For more information on observations of binary black hole mergers see Refs [S6–S11, S11, S12], for binary neutron star mergers see Refs [S11–S15] and for neutron star-black hole mergers see Refs. [S16, S17]. More information about gravitational waves can be found in Refs. [S18–S21].

### S1.3 Continuous wave sources and searches

The network of ground-based gravitational-wave observatories is also searching for continuous gravitational waves. Rotating neutron stars are prime targets for continuous wave searches. For further information about continuous waves from isolated neutron stars see for example Refs. [S22, S23] and from neutron stars in binaries see for example Ref. [S24].

The methods used in Section IV are inspired by hidden Markov model (HMM) searches for continuous gravitational waves. The HMM method is described in Refs [S25–S28] and further information can be found in Ref. [S29]. For more information on the Viterbi algorithm, the interested reader is referred to the excellent review paper by Rabiner [S30] and textbook by Quinn and Hannam [S31]. Alternative pseudocode examples (to that described in Appendix C of the main paper) can be found online at Ref. [S32].

HMM continuous-wave searches have targeted low mass X-ray binaries [S33–S35], young supernova remnants [S36–S38], post merger remnants [S39], Fomalhaut b [S40], and high energy TeV sources [S41].

A range of other continuous wave searches exist. We refer the reader to Refs. [S42–S49] for searches that consider particular targets or sky locations, and to Refs. [S49–S51] for search methods that cover the entire sky. For further information on searching for transient gravitational wave signals see for example Refs. [S52–S60].

### S1.4 Gravitational-wave engagement

Numerous outreach activities have been developed by gravitational-wave research groups around the world [S61–S72]. Specific examples of using table-top interferometers as engagement or teaching tools include Refs. [S61, S73–S76] and parts lists or instructions can be found in Refs. [S77–S81]. The analogy between sound and gravitational-wave signals is explored in Refs. [S70, S82].

## S1.5 Environmental noise in gravitational-wave detectors

In Section V B in the main article, we see evidence for environmental noise in our data. In gravitational-wave detectors reducing and mitigating environmental noise is an active area of research [S83]. One such noise source is seismic ground motion from the Earth. In gravitational-wave observatories, the optical components are suspended from a series of pendulums to achieve passive isolation, and feedback control loops provide active isolation [S3, S84]. Such a system is beyond the scope of this article. However, feedback control has been demonstrated in table-top Michelson interferometers designed for the undergraduate laboratory [S74]. Other sources of noise include acoustic sound waves (for example from nearby vehicles, wind, and components such as electric fans) and magnetic noise (for example from high-voltage power lines, building heaters and lights, or any electrical power circuit) [S83, S85]. In some cases, environmental noise sources can be removed or mitigated on site [S86]. Where this is not possible, noise subtraction can be done after data collection [S87, S88]. One example is the removal of spectral noise lines from the United States power grid operating at 60 Hz [S89].

## S1.6 Hardware and software

In Section V in the main text, we make use of a Raspberry Pi to record data from the interferometer. For further information and documentation see Refs [S90, S91].

This work is implemented in Python 3 [S92] scripts and jupyter notebooks [S93, S94] as well as in MATLAB [S95]. We make use of Python packages: numpy [S96][1], scipy [S97], matplotlib [S98], tqdm [S99], and logmmse [S100].

## S1.7 Signal processing and speech

Information on the Sallen-Key filter used in Section V.A. can be found in Ref. [S101]. We refer the reader to Ref. [S102] for details of the harmonics and non-linearity we observe in Section V.B. The cascaded notch and Wiener filter are described in Section S2 in this Supplementary Material and also in Refs. [S103, S104]. Further details about the logMMSE estimator can be found in Ref. [S105] and for speech intelligibility see Refs. [S106, S107].

# S2 Optical microphone signal processing

In Section V in the main article, we use the interferometer as an optical microphone; audio signals are used to vibrate the interferometer mirrors, the changing interference pattern is recorded, and we aim to recover the original signal from the resulting timeseries data. In this Supplementary Material, we include a selection of signal processing techniques applied to the data which may be considered for undergraduate laboratory examples. We start with some naïve approaches and then move to traditional signal processing filters (band-passing and cascaded notches). We finish by combining these with advanced statistical techniques (Wiener filter and the logMMSE estimator). As an example, we test all of these filters on the same 1 s long speech recording. For more information about the techniques used here see Refs. [S104, S108–S113].

The experimental apparatus presented here provides an opportunity to bring together physics and electrical engineering topics in the lab environment. Several examples of cross-disciplinary teaching can be found in the literature, for example, see Refs. [S114–S117]. Cross-disciplinary experimental projects between physics and engineering students can increase student understanding of overlapping concepts through contextual learning and also improve student motivation [S115]. Realistic open-ended projects spanning both topics allow students to better understand real-world applications of these topics and the interdependency between physicists and engineers (as shown by the collaborative project presented in Ref. [S114]). The optical microphone presented here can similarly be used as a teaching tool for students to conduct investigations into signal processing techniques with a physical system.

All analysis is done using the photodiode set-up as described in Section V.A. in the main article. A photograph of the equipment is shown in Fig. S1 with the interferometer and circuit shown on the left and right, respectively. The photodiode circuit diagram is shown in Figure S2. The design was based on standard examples found online of photo-detectors [S118,S119], connecting an analogy-to-digital converter ADC to a Raspberry Pi [S120], and Sallen-Key filters [S121, S122]. The Raspberry Pi documentation [S91] and the pinoutXYZ reference [S123] were also useful.

Results for a selection of filters are collated in Fig. S3, which we refer to throughout this Supplementary Material. The left and right columns show timeseries and Fourier spectrum results, respectively. The first and second rows in

[1] `https://numpy.org/doc/stable/reference/routines.fft.html`

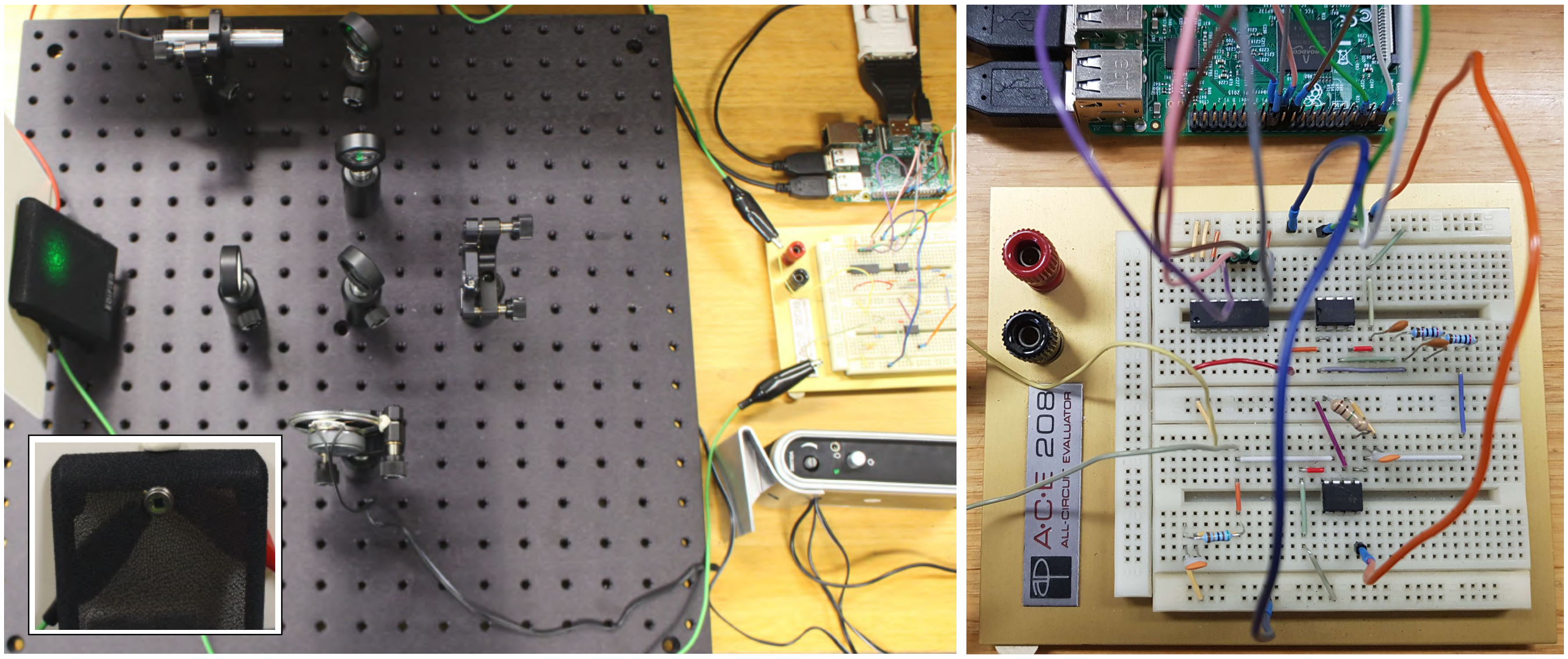

Figure S1: Left: A photograph of the optical microphone with an inset of the photodiode. The Michelson interferometer is shown on the left and the circuit and Raspberry Pi on the right. In the main image, the photodiode is placed behind a cloth screen as explained in the main article. The inset shows a face-on view of the photodiode with the cloth screen removed. Right: A photograph of the photodiode circuit assembled on a breadboard. The leads from the photodiode enter from the left of the picture. The Raspberry Pi is shown at the top of the picture.

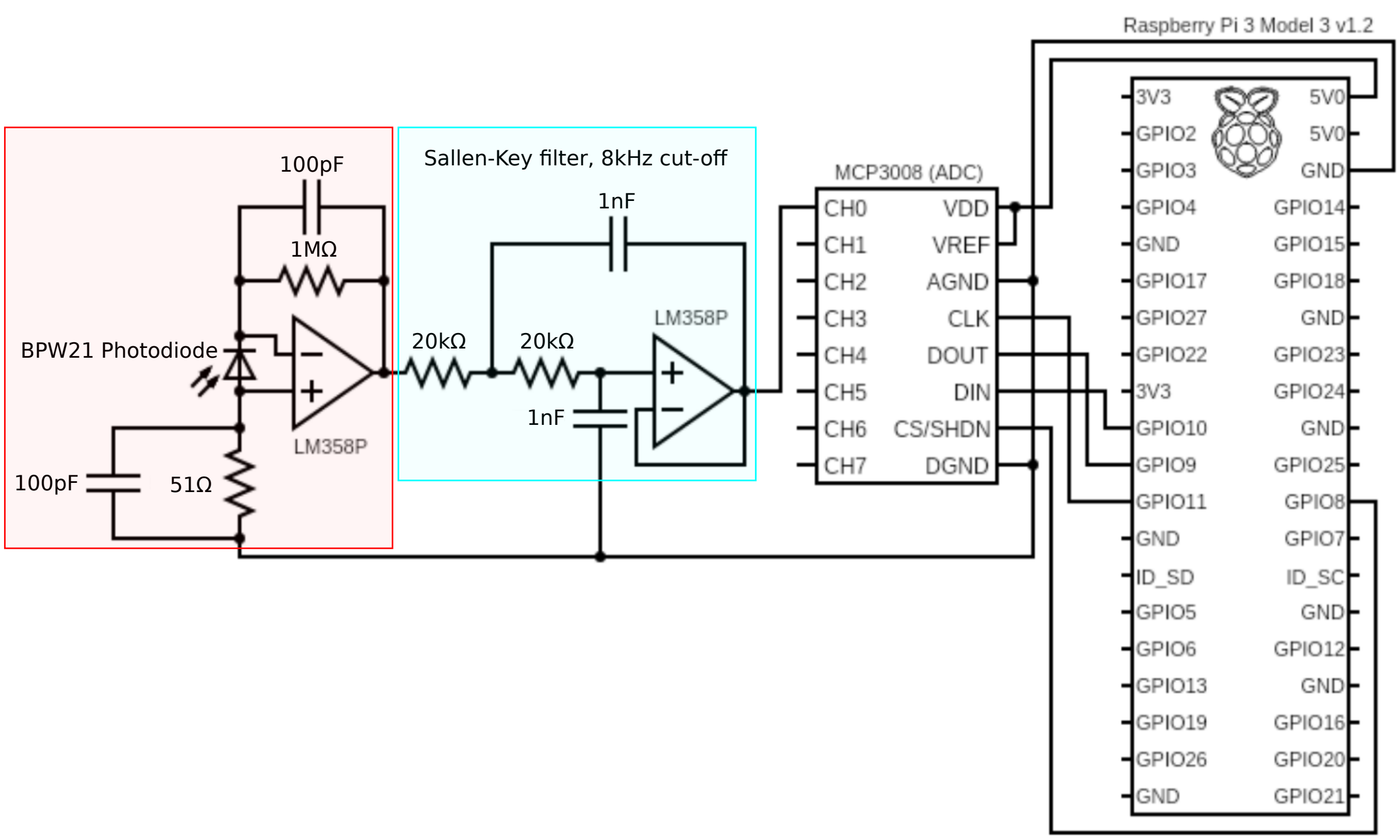

Figure S2: Circuit diagram for reading the photodiode. The photodiode is connected in reverse-bias across an op-amp (pale-red box). The op-amp output is passed through a Sallen-Key anti-aliasing filter (pale-blue box) with a cut-off frequency of 8 kHz, then into an analog-to-digital converter (ADC). The digitized signal is then read by the special-purpose input (SPI) pins of a Raspberry Pi in standard configuration. This diagram was produced using the Circuit Diagram online tool [S124].

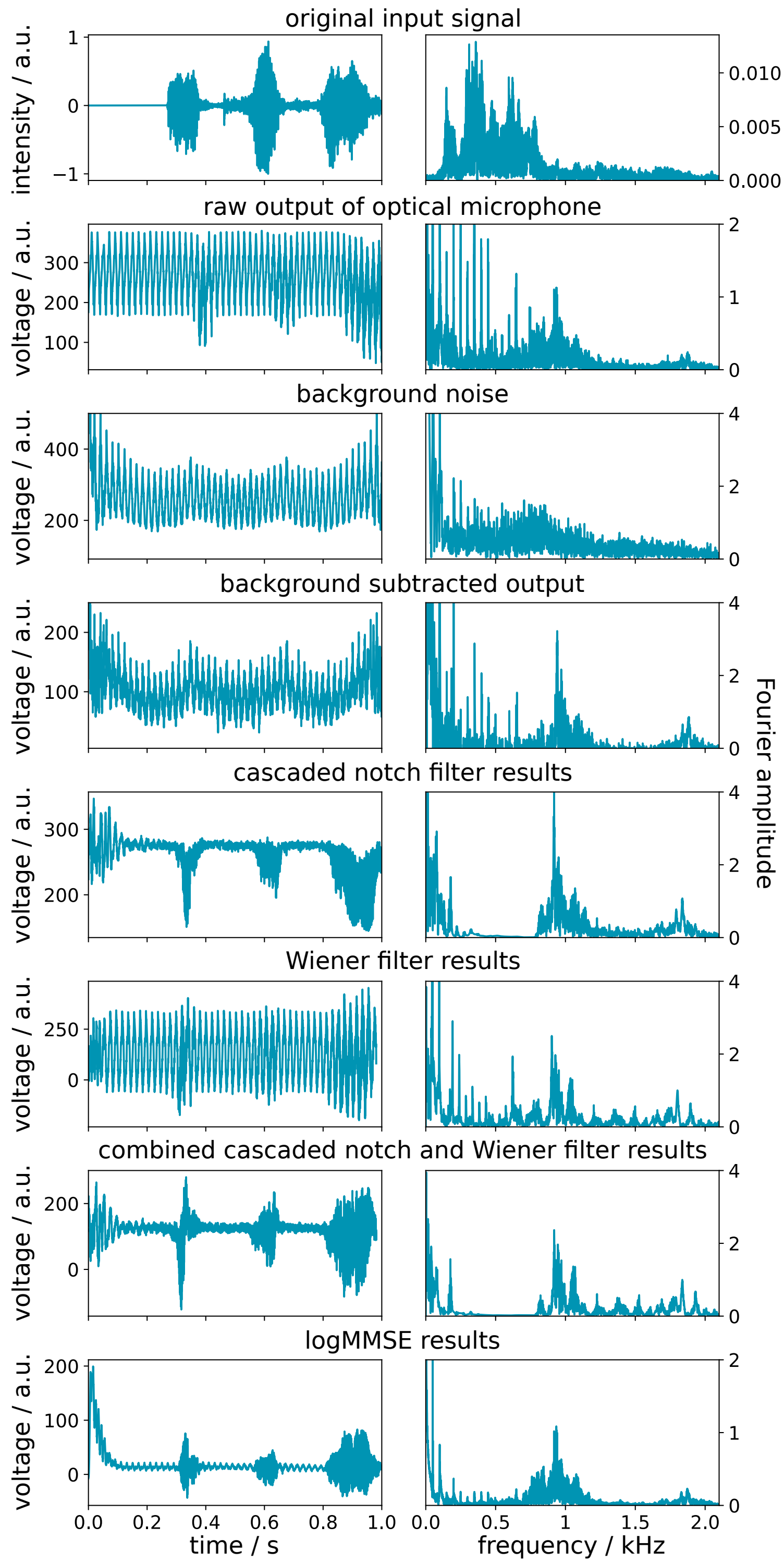

Figure S3: Timeseries (left column) and frequency spectrum (right column) results for the background subtraction, notch, and Wiener filters. The top two rows are the input signal and raw output (identical to the top two rows in Fig. 7 in the main article). The third row shows a background noise recording from the optical microphone when no audio signal is played. The fourth row shows the result of subtracting the background noise spectrum in the third row from the recording in the second row. The fifth and sixth rows show the results of applying the cascaded-notch filter and Wiener filter respectively. The seventh shows the results of applying both the cascaded-notch and Wiener filter (identical to the third row in Fig. 7 in the main article). The eighth row shows the results of applying the logMMSE estimator (identical to the fourth row in Fig. 7 in the main article).

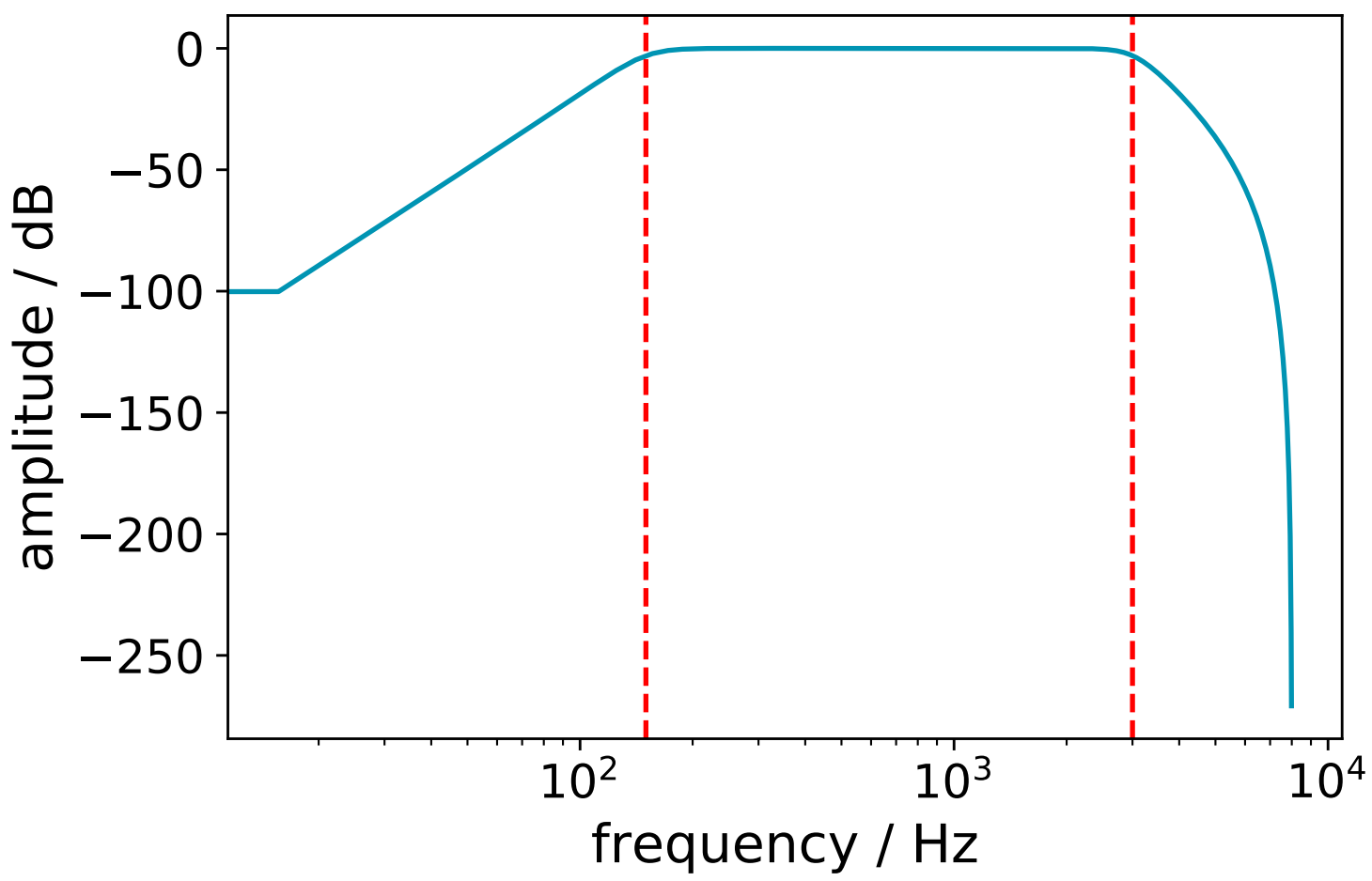

Figure S4: Butterworth bandpass filter frequency response. The red, dashed vertical lines show the band limits of 150 Hz and 3 kHz. Note the flat response within the band characteristic of the Butterworth filter.

Fig. S3 show the original, source signal and the raw optical microphone recording, respectively. The following sections describe a sequence of analysis techniques and a selection of their results is presented in Fig. S3.

## S2.1 Background noise subtraction

The first technique we consider is quite intuitive: we try to remove noise by directly subtracting the background noise from the recorded spectrum. The third row in Fig. S3 shows the interferometer background noise when there is no input signal. The fourth row shows the spectrum obtained after subtracting the background noise spectrum, i.e. the result of subtracting the third row from the second row. We see no obvious improvement, which may be attributed to a time-variant noise spectrum, the cause of which is unidentified.

## S2.2 Rectangular comb filter

Since the mains harmonics are present in the signal, we try to selectively remove them. However, simply zeroing the frequency bins corresponding to the harmonics of the 50 Hz mains signal is unsuccessful. This effectively multiplies the spectrum by a rectangular comb filter. It does remove the mains harmonics, but audibly ruins the rest of the signal due to lack of smoothness. This is because applying a filter in frequency space is equivalent to convolving the time-domain signal with the inverse Fourier transform of that filter. The inverse Fourier transform of a rectangular comb filter (a set of boxcars) is some combination of sinc functions, which significantly corrupt the signal. See also Section S2.5 where we explore a notch filter.

## S2.3 High-pass filter

A high-pass filter smoothly attenuates frequencies below some cut-off frequency. Applying a high-pass filter, with a cut-off frequency around 150 Hz, to the signal spectrum works well at removing the 50 Hz and 100 Hz harmonics. However, the mains harmonics above 100 Hz remain. A high-pass filter with a higher cut-off can be used to mitigate this issue. However, it makes the played-back signal unrecognizable as the region above 100 Hz carries a lot of the fundamental frequencies of speech and music [S106]. Often in speech processing, the logarithm of the signal is taken since the amplitude information seems to be more important to intelligibility than the phase information to the human ear [S107]. However, applying a high-pass filter to the logarithm of the signal spectrum does not significantly improve on the above simple high-pass filter.

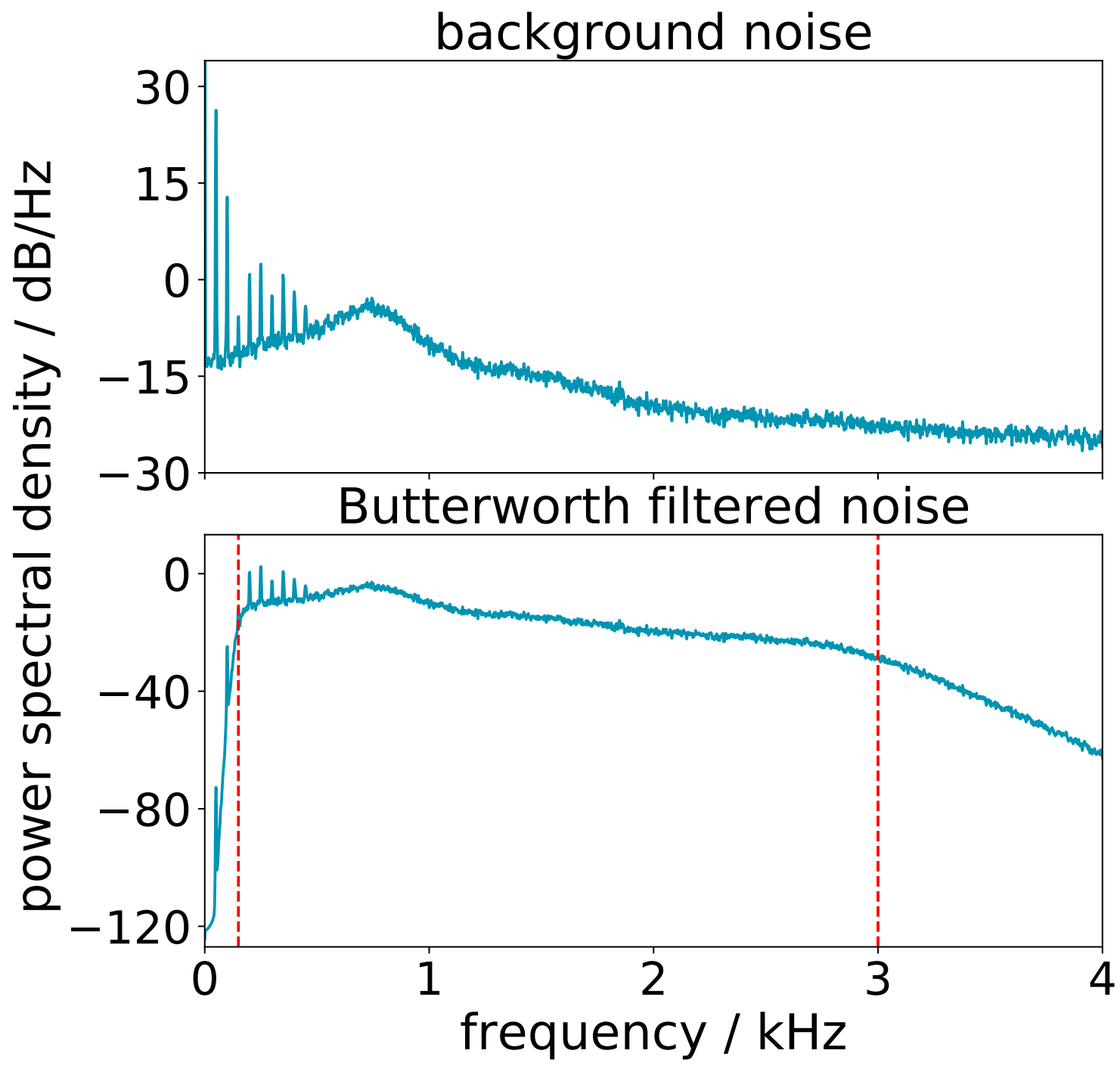

Figure S5: Top: power spectral density (PSD) of background noise from the optical microphone (with the speaker off). Bottom: the PSD after applying a Butterworth bandpass filter (bottom panel) between the two frequencies marked with red, dashed lines. We see strong power from the 50 Hz mains hum and its harmonics (most likely from the photodiode circuit and the room's lighting and cooling). Otherwise, the PSD is fairly white. After filtering, we see significant attenuation (at least 3 dB) of all frequencies outside the band. However, the filter has little change on the harmonics within the band. The sharp drop-off at low frequencies, around 100 dB near 0 Hz, is due to stronger attenuation far from the band and would appear shallower if shown against a logarithmic scale for frequency as in Fig. S4.

### S2.4 Butterworth band-pass filter

A general band-pass filter combines a high-pass filter and a low-pass filter to smoothly attenuate frequencies outside of some band (alt. pass-band). A Butterworth band-pass filter is a particular band-pass filter such that the frequency response (the attenuation at each frequency) is "maximally flat" within the band. The Butterworth low-pass component is given by

$$H(f) = \left[1 + \varepsilon^2 \left(\frac{f}{f_c}\right)^{2n}\right]^{-1/2}, \tag{S1}$$

and is combined with a similar high-pass filter to form the band-pass filter. In Eq. S1, $f_c$ is the cut-off frequency of the low-pass Butterworth filter, $\varepsilon$ is the gain, and $n$ is the order of the filter which determines how quickly the response rolls off past the cut-off frequency. Fig. S4 shows the frequency response of the filter used here: a fifth order ($n = 5$) Butterworth filter with a pass-band of (150 Hz, 3 kHz). This high frequency (3 kHz) cut-off is chosen since the frequencies important for speech generally lie below 2 kHz [S106].

The effect of applying this filter to the background noise PSD can be seen in the bottom panel in Fig. S5. The Butterworth band-pass filter reduces the amplitude of mains harmonics below 150 Hz and suppresses unrelated noise sources above 3 kHz. However, it does not address the issue of mains harmonics above 150 Hz (i.e. those in the pass-band). In the following section, we experiment with a cascade notch filter to address this.

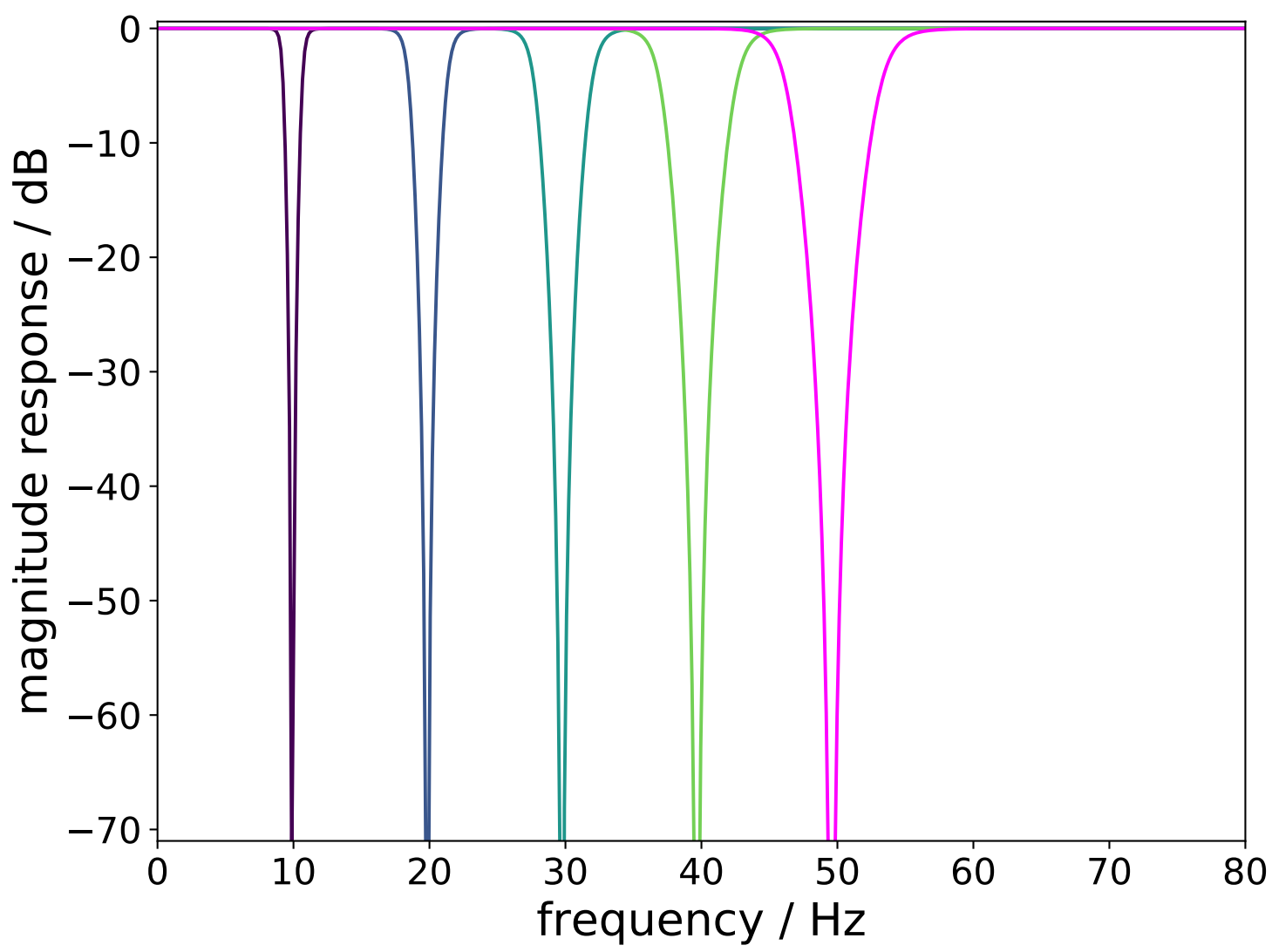

Figure S6: Amplitude (magnitude) response for the first five notches of fifteen for the cascaded notch filter described by Eqs. S2 and S3.

## S2.5 Cascade notch filter

One method to remove mains noise and its harmonics is to use a sequence of notch filters centered at each of the frequency bins we want to remove. This sequence is known as a "cascade" of notches. Here, the notches are smooth in comparison to the naïve zeroing of each frequency (which looks like a rectangular comb).

The impulse response describes the reaction of a system to an input signal as a function of time. If a system is linear and time-invariant, the output signals are completely determined for any input signal using the impulse response. Finite impulse response (FIRs) filters have finite length (they equal zero outside of a finite range). In contrast, an infinite impulse response (IIR) filter has infinite length due to feedback. Typically FIR filters outperform IIR filters as they are always stable. However, IIR filters normally require fewer coefficients which in turn speeds up signal processing in comparison to FIR filters.

In this work, we opt for an IIR notch filter. We write the filter in the form of a Z-transform, in which a discrete-time signal is converted to the complex frequency domain $z = e^{jf}$. The IIR notch filter we use here is given by [S103],

$$H(z) = \frac{1+\alpha}{2} \frac{1 - 2\beta z^{-1} + z^{-2}}{1 - \beta(1+\alpha)z^{-1} + \alpha z^{-2}} \,, \tag{S2}$$

where $\alpha$ and $\beta$ are parameters that control the filter. These parameters determine $w_0 = \cos^{-1}(\beta)$, the frequency that is completely attenuated (zeroed or "notched") at the centre of the notch, and $B_w = \cos^{-1}[2\alpha/(1+\alpha^2)]$, the bandwidth of the notch which determines how quickly the response changes around the notched frequency. We find that a sequence of 15 notches works well here, with the $k^{\text{th}}$ notch centered on the $k^{\text{th}}$ harmonic of the 50 Hz mains signal, where $k = 0, 2, \ldots, 14$. We choose the bandwidth and order (here order six) of each notch to avoid disturbing useful signals while still allowing for uncertainty in the location of each harmonic of the mains signal. The response of this cascaded notch filter $H(z)$ is the product of the responses of each of the individual $H_k(z)$ notches,

$$H(z) = \prod_{k=0}^{14} H_k(z). \tag{S3}$$

We use the built-in MATLAB filter design toolbox in this work [S95]. The amplitude (magnitude) response of the first five notch filters is shown in Fig. S6. The time series and spectrum obtained after applying the cascade notch filter to the speech recording are shown in the fifth row in Fig. S3. We see that the mains harmonics are significantly attenuated compared to the background-subtracted results shown in the fourth row in Fig. S3.

Although the notch filter removes much of the mains hum sound, the filtered recording is not intelligible. Qualitatively, it sounds more like a drum than a human voice. This is due to the loss of voice information under the filter. To overcome this, we turn to more advanced techniques, starting with the Wiener filter. Instead of just passively filtering different frequencies, this statistical technique optimizes an estimate of the original signal.

## S2.6 Wiener filter

A Wiener filter is an advanced statistical technique that estimates the injected signal given prior information about the injected spectrum and the reference spectrum of the background noise. The observed noisy speech signal sequence is given as $\mathbf{x} = (x(0), \ldots, x(N-1))$, where $N$ is the length of the data sequence and $\mathbf{x}$ is the sum of the original injected signal $\mathbf{s} = (s(0), \ldots, s(N-1))$, and the noise sequence $\mathbf{w} = (w(0), \ldots, w(N-1))$,

$$\mathbf{x} = \mathbf{s} + \mathbf{w}\,. \tag{S4}$$

Given $\mathbf{x}$, our goal is to make an estimate $\hat{\mathbf{s}}$ of the original signal $\mathbf{s}$ such that we minimise the Bayesian mean-square error (BMSE) between the two, defined as

$$\text{BMSE}(\hat{\mathbf{s}}) = E[(\mathbf{s} - \hat{\mathbf{s}})^2]\,. \tag{S5}$$

If we assume that: (i) $\mathbf{x}$ is "wide sense stationary" [2] with zero mean; (ii) the signal $\mathbf{s}$ has a mean of zero; and (iii) the noise $\mathbf{w}$ is uncorrelated with the signal $\mathbf{s}$, we can further express $\hat{\mathbf{s}}$ to be a linear combination of present and past observed data

$$\hat{s}[n] = \sum_{k=0}^{n} h[k]x[n-k]\,, \tag{S6}$$

where $\mathbf{h} = (h(0), \ldots h(n))$ represents the coefficients of an $n$th order Wiener filter. The famous Wiener-Hopf equation [S125] allows us to determine $\mathbf{h}$, as

$$\begin{bmatrix} r_{xx}[0] & r_{xx}[1] & \ldots & r_{xx}[n] \\ r_{xx}[1] & r_{xx}[0] & \ldots & r_{xx}[n-1] \\ \vdots & \vdots & \ddots & \vdots \\ r_{xx}[n] & r_{xx}[n-1] & \ldots & r_{xx}[0] \end{bmatrix} \begin{bmatrix} h[0] \\ h[1] \\ \vdots \\ h[n] \end{bmatrix} = \begin{bmatrix} r_{ss}[0] \\ r_{ss}[1] \\ \vdots \\ r_{ss}[n] \end{bmatrix}, \tag{S7}$$

where $r_{xx}$ and $r_{ss}$ are the auto-correlation functions of $\mathbf{x}$ and $\mathbf{s}$ between timestep $i$ and $i+n$,

$$r_{xx}[n] = E[x(i)\;x(i+n)]\,, \tag{S8}$$
$$r_{ss}[n] = E[s(i)\;s(i+n)]\,. \tag{S9}$$

If we let the Wiener filter be non-causal (i.e. we estimate the current signal based on both past *and future* observations), then we can represent Eq. S7 in the frequency domain as

$$H(f) = \frac{P_{ss}(f)}{P_{xx}(f)} = \frac{P_{ss}(f)}{P_{ss}(f) + P_{ww}(f)}\,, \tag{S10}$$

where $P_{xx}(f), P_{ss}(f), P_{ww}(f)$ are the spectra of the observed noisy data, the injected signal, and the background noise, respectively. Intuitively, from Eq. S10, we can see that the non-causal Wiener filter amplifies the input signal where the signal-to-noise ratio (SNR) is high and attenuates the signal where the SNR is low. The causal Wiener filter (which only makes estimates based on past observations) is similar. A more detailed analysis of both kinds of Wiener filters can be found in Ref. [S104].

In this work, we construct a higher-order causal Wiener filter based on Eq. S7. A higher-order Wiener filter provides greater smoothing of the input signal but also increases the computational memory required. For this work, we choose a Wiener filter of order $n = 100$ as it provides a reasonable balance between smoothing and efficiency. The timeseries and frequency spectrum after applying the Wiener filter are shown in the sixth row in Fig. S3. We see a significant improvement in the timeseries of the recovered signal, however, a strong noise hum persists, audibly.

[2] A random process $\{x(t)\}$ is wide sense stationary if, for all $t_1, t_2 \in \mathbb{R}$, (1) its mean is time invariant, i.e., $\mu_x(t_1) = \mu_x(t_2) =$ constant; and (2) the autocorrelation depends only on the time difference, i.e., $R_x(t_1, t_2) = R_x(\tau), \tau = t_1 - t_2$.

### S2.7 Combined notch and Wiener filter

Here, we experiment with applying a combination of the cascaded notch and the Wiener filter to the recorded speech signal. The results of this analysis also are described in Section V.C. in the main article.

The Wiener filter makes use of statistical information from the speech data and noise. It amplifies the part of the signal with high SNR while suppressing the parts with low SNR (see above Sec. S2.6). It is implemented in the form of a finite response filter, which ensures linear phase response and stability (both desirable), but at the cost of high orders computationally. By comparison, the notch filter is based on directly removing the unwanted frequency components. It is implemented in the form of an infinite impulse response filter. Although it significantly decreases the order of the overall filter, it unavoidably introduces nonlinear phase and instability.

By combining the notch and Wiener filter, we can trade-off between the two and achieve an overall better performance, as can be seen in the seventh row in Fig. S3. The filtered voice after the combined notch and Wiener filter is enhanced compared to either alone. The mains noise is all but removed and more voice information is retained. However, the recovered voice still sounds muffled and is not understandable.

### S2.8 logMMSE estimator

Speech enhancement of noisy channels is a classic problem in signal processing. In Ref. [S107], a comparison is made of 13 speech enhancement methods, finding the log minimum mean-square error (logMMSE) estimator to be the best, qualitatively, at recovering speech. This estimator is based on speech enhancement techniques discussed in Ref. [S105] and minimizes the mean-square error (MSE) of the estimate from the injected signal, like the Wiener filter above, except that it measures the MSE between the logarithm of the Fourier amplitudes. This is motivated by the fact that the logarithm approximates the response of the human ear [S107]. We apply an existing implementation of the logMMSE estimator (see Ref. [S100]) to the recorded signal.

The final row in Fig. S3 shows the results of the logMMSE estimator. Again there is significant noise reduction, however, the speech remains indistinct. The logMMSE results are also presented in Section V.C. in the main article.